\documentclass[reqno]{amsart}

\usepackage{amsmath,amsfonts}
\usepackage{epsfig}
\usepackage{graphicx}

\def\alphaset{{\mathfrak A}}

\def\diffN{\chi_N}
\def\duh{{\rm Duh}}

\def\gammaN{\gamma_N}

\def\GammaN{\Gamma_N}

\def\Gspace{{\mathfrak G}}

\def\opDelta{\widehat{\Delta}}
\def\opB{B}
\def\opBN{\opB_N}
\def\opU{\widehat{U}}

\def\VN{V_N}

\def\supp{{\rm supp}}  

\def\tr{{\rm Tr}}

\def\bra{\big\langle}
\def\ket{\big\rangle}
\def\Bra{\Big\langle}
\def\Ket{\Big\rangle}

\def\tq{\widetilde{q}}

\def\tx{\widetilde{x}}
\def\tux{\widetilde{\ux}}

\def\tu{\widetilde{u}}

\def\C{{\mathbb C}}
\def\N{{\mathbb N}}

\def\R{{\mathbb R}}

\def\uu{{\underline{u}}}

\def\ux{{\underline{x}}}

\def\frh{{\mathfrak h}}

\def\cD{{\mathcal D}}
\def\cH{{\mathcal H}}

\def\cM{{\mathcal M}}

\def\cP{{\mathcal P}}

\def\1{{\bf 1}}

\def\eqnn{\begin{eqnarray*}}
\def\eeqnn{\end{eqnarray*}}
\def\eqn{\begin{eqnarray}}
\def\eeqn{\end{eqnarray}}

\def\prf{\begin{proof}}
\def\endprf{\end{proof}}

\theoremstyle{plain}
\newtheorem{theorem}{Theorem}[section]
\newtheorem{definition}[theorem]{Definition}
\newtheorem{proposition}[theorem]{Proposition}

\newtheorem{lemma}[theorem]{Lemma}

\newtheorem{remark}[theorem]{Remark}

\numberwithin{equation}{section}

\begin{document}

\parskip=8pt

\title[Derivation of the cubic GP hierarchy]
{Derivation of the cubic NLS and  Gross-Pitaevskii hierarchy  
from manybody dynamics
in $d=3$ based on  spacetime norms}

\author[T. Chen]{Thomas Chen}
\address{T. Chen,  
Department of Mathematics, University of Texas at Austin.}
\email{tc@math.utexas.edu}

\author[N. Pavlovi\'{c}]{Nata\v{s}a Pavlovi\'{c}}
\address{N. Pavlovi\'{c},  
Department of Mathematics, University of Texas at Austin.}
\email{natasa@math.utexas.edu}

%\date{\mscrptdate}

\begin{abstract}
We derive the defocusing cubic Gross-Pitaevskii (GP) hierarchy in
dimension $d=3$, from an $N$-body Schr\"{o}dinger equation 
describing a gas of interacting bosons in the GP scaling, in the limit $N\rightarrow\infty$. 
The main result of this paper is the proof of convergence  
of the corresponding BBGKY hierarchy to a GP hierarchy
in the spaces introduced in our previous work on the well-posedness 
of the Cauchy problem for GP hierarchies, \cite{chpa2,chpa3,chpa4}, which are inspired by
the solutions spaces based on space-time norms 
introduced by Klainerman and Machedon in \cite{klma}. 
We note that in $d=3$, this has been a well-known open problem in the field.
While our results do not assume factorization of the solutions, consideration
of factorized solutions yields
a new derivation of the cubic, defocusing
nonlinear Schr\"odinger equation (NLS) in $d=3$.
\end{abstract}

\maketitle

\section{Introduction}
\label{sec-intro}

We derive the defocusing cubic 
Gross-Pitaevskii (GP) hierarchy from an $N$-body Schr\"{o}dinger equation 
in dimension $d=3$
describing a gas of interacting bosons in the Gross-Pitaevskii (GP) scaling,
as $N\rightarrow\infty$.
The main result of this paper is the proof of convergence  
in the spaces introduced in our previous work on the well-posedness 
of the Cauchy problem for GP hierarchies, \cite{chpa2,chpa3,chpa4}, which are inspired by
the solutions spaces based on space-time norms 
introduced by Klainerman and Machedon in \cite{klma}.
In dimension 3, this problem has so far remained a key open problem,
while in dimensions 1 and 2, it was solved in \cite{kiscst,chpa} for the
cubic and quintic case.

The derivation of nonlinear dispersive PDEs,
such as the nonlinear Schr\"odinger (NLS) or nonlinear Hartree (NLH) equations, 
from many body quantum dynamics is a central topic in mathematical physics,
and has been approached by many authors in a variety of ways;
see  \cite{esy1,esy2,ey,kiscst,klma,rosc} and the references therein,
and also \cite{adgote,anasig,xch1,xch2,eesy,frgrsc,frknpi,frknsc,grma,grmama,he,pick,sp}.
This problem is closely related to the phenomenon of Bose-Einstein condensation (BEC)
in systems of interacting bosons, which was first experimentally verified in 1995,
\cite{anenmawico,dameandrdukuke}. For the mathematical study of BEC, we refer to the the 
fundamental works
\cite{ailisesoyn,lise,lisesoyn,liseyn} and the references therein.

\subsection{The Gross-Pitaevkii limit for Bose gases}  
\label{subsec-exsol-GP-0}

As a preparation for our analysis in the present paper, we will
outline some main ingredients of the approach due to L. Erd\"os, B. Schlein, and H.-T. Yau.
In an important series of works, \cite{esy1,esy2,ey}, these authors
developed a powerful method to derive the cubic nonlinear Schr\"odiner equation
(NLS) from 
the dynamics of an interacting Bose gas in the Gross-Pitaevskii limit. 
We remark that the defocusing quintic NLS can be derived from a system of
bosons with repelling three body interactions, see \cite{chpa}.   

\subsubsection{From $N$-body Schr\"odinger to BBGKY hierarchy}
We consider a quantum mechanical system consisting of $N$ bosons in $\R^3$ 
with wave function 
$\Phi_{N} \in L^2(\R^{3N})$. 
According to Bose-Einstein statistics,  
$\Phi_{N}$ is invariant under the permutation of particle variables,  
\begin{equation}\label{sym}
	\Phi_N(x_{\pi( 1)},x_{\pi (2)},...,x_{\pi (N)}) \, = \, \Phi_N(x_1, x_2,..., x_N) \
	\; \; \; \; \; \; \; \; \forall \pi \in S_{N} \,,
\end{equation} 
where  $S_{N}$ is the $N$-th symmetric group.
We denote by $L_{sym}^{2}(\mathbb R^{3N})$ the subspace  of 
$L^2(\R^{3N})$ of elements obeying \eqref{sym}. 
The dynamics of the system is determined by the $N$-body Schr\"odinger equation
\begin{equation}\label{ham1}
i\partial_{t}\Phi_{N} \, = \, H_{N}\Phi_{N} \,.
\end{equation} 
The Hamiltonian $H_{N}$ is given by a self-adjoint operator acting on the 
Hilbert space $L_{sym}^2(\mathbb R^{3N})$, of the form
\begin{equation}\label{ham2}
	H_{N} \, = \, \sum_{j=1}^{N}(-\Delta_{x_{j}})+\frac1N\sum_{1\leq i<j \leq N}V_N(x_{i}-x_{j}),
\end{equation}
where $V_N(x)=N^{3\beta}V(N^\beta x)$ with
$V\geq0$ spherically symmetric, sufficiently regular, 
and for $0<\beta<\frac1{4}$. 

Since the Schr\"{o}dinger equation \eqref{ham1} is linear and $H_N$ self-adjoint,  the global 
well-posedness of solutions is evident.  
To perform the infinite particle number limit $N \rightarrow \infty$,
we outline the strategy developed in \cite{esy1,esy2}  as follows. 

One introduces the density matrix
$$\gamma_{\Phi_N}(t, \ux_N, \ux_N')= |\Phi_{N}(t,\ux_N) \, \rangle\langle \, \Phi_{N}(t,\ux'_N)| :=
\Phi_{N}(t, \ux_N)\overline{\Phi_{N}(t,\ux_N')}$$
where $\ux_N=(x_1,x_2,..., x_N)$ and $\ux_N'=(x_{1}^{\prime}, x_{2}^{\prime},..., x_{N}^{\prime})$.
Furthermore, one considers the associated sequence of 
$k$-particle marginal density matrices  $\gamma_{\Phi_N}^{(k)}(t)$, for $k=1,\dots,N$, 
as the partial traces of $\gamma_{\Phi_N}$ 
over the degrees of freedom associated to the last $(N-k)$ particle variables,
$$\gamma_{\Phi_N}^{(k)}=\tr_{k+1, k+2,...,N}|\Phi_{N}\rangle\langle\Phi_{N}| \,.$$
Here, $\tr_{k+1, k+2,...,N}$ denotes the partial trace with respect to the particles indexed
by $k+1, k+2,..., N$. 
Accordingly, $\gamma_{\Phi_N}^{(k)}$ is explicitly given by  
\eqn
	\gamma_{\Phi_N}^{(k)}(\ux_k,\ux_k')
	&=& \int d\ux_{N-k} \gamma_{\Phi_N}(\ux_k, \ux_{N-k};\ux_k', \ux_{N-k})
	\nonumber\\
	\label{reduced}
	&=&\int d\ux_{N-k} \Phi_{N}(\ux_k, \ux_{N-k}) \overline{ \Phi_{N}(\ux_k', \ux_{N-k}) } \,.
\eeqn
It follows immediately from the definitions that the property of {\em admissibility} holds,
\eqn
	\gamma^{(k)}_{\Phi_N} \, = \, \tr_{k+1}(\gamma^{(k+1)}_{\Phi_N})
	\; \; \; \; \;  , \; \; \; \; k\, = \, 1,\dots,N-1 \,,
\eeqn 
for $1\leq k\leq N-1$, and that 
$\tr \gamma_{\Phi_N}^{(k)}=\|\Phi_N\|_{L_{s}^{2}(\mathbb R^{3N})}^2=1$ for all $N$, 
and all $k=1, 2, ..., N$. 

Moreover, $\gamma_{\Phi_N}^{(k)}\geq0$ is positive semidefinite as an operator 
${\mathcal S}(\R^{3k})\times{\mathcal S}(\R^{3k})\rightarrow\C$, 
$(f,g)\mapsto \int d\ux d\ux'f(\ux)\gamma(\ux;\ux')\overline{g(\ux')}$.

The time evolution of the density matrix $\gamma_{\Phi_N}$ is determined by the 
Heisenberg equation
\begin{equation}\label{von}
	i\partial_{t}\gamma_{\Phi_N}(t) \, = \, [H_{N}, \gamma_{\Phi_N}(t)] \, ,
\end{equation}
which has the explicit form
\eqn
	i\partial_{t}\gamma_{\Phi_N}(t,\ux_N,\ux_N')
	&=&-(\Delta_{\ux_N}-\Delta_{ \ux_N' })\gamma_{\Phi_N}(t,\ux_N, \ux_N') 
	\label{von2}\\
	&&+\frac{1}{N}\sum_{1\leq i<j \leq N}[V_N(x_i-x_j)-V_N(x_i^{\prime}-x_{j}^{\prime})]
	\gamma_{\Phi_N}(t, \ux_N,\ux_N') \,.
	\nonumber
\eeqn 
Accordingly, the $k$-particle marginals satisfy the BBGKY hierarchy 
\eqn\label{BBGKY}
	\lefteqn{
	i\partial_{t}\gamma_{\Phi_N}^{(k)}(t,\ux_k;\ux_k')
	 \, = \, 
	-(\Delta_{\ux_k}-\Delta_{ \ux_k'})\gamma_{\Phi_N}^{(k)}(t,\ux_k,\ux_k')
	}
	\nonumber\\
	&&
	+ \frac{1}{N}\sum_{1\leq i<j \leq k}[V_N(x_i-x_j)-V_N(x_i^{\prime}-x_{j}')]
	\gamma_{\Phi_N}^{(k)}(t, \ux_{k};\ux_{k}') 
	\label{eq-bbgky-1}\\
	&&+\frac{N-k}{N}\sum_{i=1}^{k}\int dx_{k+1}[V_N(x_i-x_{k+1})-V_N(x_i^{\prime}-x_{k+1})]
	\label{eq-bbgky-2}\\
	&&\quad\quad\quad\quad\quad\quad\quad\quad\quad\quad\quad\quad\quad\quad\quad\quad
	\gamma_{\Phi_N}^{(k+1)}(t, \ux_{k},x_{k+1};\ux_{k}',x_{k+1})
	\nonumber
\eeqn 
where $\Delta_{\ux_k}:=\sum_{j=1}^{k}\Delta_{x_j}$, and similarly for $\Delta_{\ux_k'}$. 
We note that the number of terms in (\ref{eq-bbgky-1}) is $\approx \frac{k^2}{N}\rightarrow0$, 
and the number of terms in (\ref{eq-bbgky-2}) is $\frac{k(N-k)}{N}\rightarrow k$ 
as $N\rightarrow \infty$. Accordingly, for fixed $k$, (\ref{eq-bbgky-1}) disappears in the limit 
$N\rightarrow\infty$ described below, while (\ref{eq-bbgky-2}) survives.
 
\subsubsection{From BBGKY hierarchy to GP hierarchy.} 
It is proven in \cite{esy1,esy2,ey} that, for asymptotically factorized initial data,
and in the weak topology on the space of marginal
density matrices, one can extract convergent subsequences
$\gamma^{(k)}_{\Phi_N}\rightarrow\gamma^{(k)}$ 
as $N\rightarrow\infty$, for   $k\in\N$, which satisfy the  
the infinite limiting hierarchy  
\eqn\label{eq-GP-0-0}
	i\partial_{t}\gamma^{(k)}(t,\ux_k;\ux_k')
	&=&
 	- \, (\Delta_{\ux_k}-\Delta_{\ux_k'})\gamma^{(k)}(t,\ux_k;\ux_k')
	\label{GP}\\
	&&+ \, \kappa_0 \sum_{j=1}^{k}   \left(B_{j, k+1} 
	\gamma^{k+1}\right)(t,\ux_k  ; \ux_k' )  \,, 
	\nonumber
\eeqn
which is referred to as the {\em Gross-Pitaevskii (GP) hierarchy}.
Here,
\eqn
	\lefteqn{
	(B_{j, k+1} \gamma^{k+1})(t,  \ux_k ; \ux_k' )
	}
	\nonumber\\
	&:=&
	\int dx_{k+1}dx_{k+1}'[\delta(x_j-x_{k+1})\delta(x_{j}-x_{k+1}^{\prime})-\delta(x_j^{\prime}-x_{k+1})
	\delta(x_{j}^{\prime}-x_{k+1}^{\prime})]
	\nonumber\\
	&&\quad\quad\quad\quad\quad\quad\quad\quad\quad\quad\quad\quad\quad\quad\quad\quad
	\gamma^{(k+1)}(t,\ux_k, x_{k+1};\ux_k', x'_{k+1}) \,.
	\nonumber 
\eeqn
The coefficient  $\kappa_0$ is the {\em scattering length} if $\beta=1$ (see \cite{esy1,lisesoyn} for the definition),
and $\kappa_0=\int V(x) dx$ if 
$\beta<1$ (corresponding to the Born approximation of the scattering length).
For $\beta<1$, the interaction term is obtained from the weak limit $V_N(x)\rightarrow \kappa_0\delta(x)$ 
in  (\ref{eq-bbgky-2}) as 
$N\rightarrow\infty$. The proof for the case $\beta=1$ is much more difficult, and the derivation
of the scattering length in this context is a breakthrough result obtained in \cite{esy1,esy2}.
For notational convenience, we will mostly set $\kappa_0=1$ in the sequel.

Some key properties satisfied by the solutions of the GP hierarchy are:
\begin{itemize}
\item
The solution of the GP hierarchy obtained in \cite{esy1,esy2} exists {\em globally} in $t$.
\item
It satisfies the property of admissibility,
\eqn
	\gamma^{(k)} \, = \, \tr_{k+1}(\gamma^{(k+1)})
	\; \; \; \; , \; \; \; \; 
	\forall \; k\in\N \,,
\eeqn 
which is inherited from the system at finite $N$.
\item
There exists a constant $b_1'$ depending on the initial data only, such that
the {\em a priori energy bound}
\eqn\label{eq-ESY-apriori-enbd-0}
	\tr( \, | S^{(k,1)}\gamma^{(k)}(t)| \, ) \, < \, (b_1')^k
\eeqn
is satisfied for all $k\in \N$, and for all $t\in\R$, where
\eqn
	S^{(k,\alpha)} \, := \, 
	\prod_{j=1}^k \langle\nabla_{x_j}\rangle^\alpha\langle\nabla_{x_j'}\rangle^\alpha \,.
\eeqn
This is obtained from energy conservation in the original $N$-body Schr\"{o}dinger system.
\item
Solutions of the GP hierarchy are studied in   spaces
of $k$-particle marginals $\{\gamma^{(k)} \, | \, \|\gamma^{(k)}\|_{\frh^1}  \, < \, \infty\}$ with norms 
\eqn\label{eq-frhnorm-def-0}
	\|\gamma^{(k)}\|_{\frh^{\alpha}} \, := \,  \tr (|S^{(k,\alpha)}\gamma^{(k)}|)   \,.
\eeqn
This is in agreement with the a priori bounds \eqref{eq-ESY-apriori-enbd-0}.
\end{itemize}

\subsubsection{Factorized solutions of GP and NLS}

The NLS emerges as the mean field dynamics of the Bose gas for the very special
subclass of solutions of the GP hierarchy that are {\em factorized}.  
Factorized  $k-$particle marginals at time $t=0$ have the form
$$\gamma_0^{(k)}(\ux_k;\ux_k')=\prod_{j=1}^{k}\phi_0( x_{j})\overline{\phi_0( x_{j}^{\prime}})\,,$$
where we assume that $\phi_0\in H^1(\R^3)$. One can easily verify that 
$$\gamma^{(k)}(t,\ux_k;\ux_k')=\prod_{j=1}^{k}\phi(t,x_{j})\overline{\phi(t,x_{j}^{\prime})} \,,$$
is a solution  (usually referred to as a factorized solution)
of the  GP hierarchy \eqref{eq-GP-0-0} with
$\kappa_0=1$,  if $\phi(t)\in H^1(\R^3)$ solves the defocusing cubic NLS,
\eqn
	i\partial_t\phi \, = \, - \Delta_x \phi \, + \, |\phi|^2\phi\,,
\eeqn 
for $t\in I\subseteq\R$, and $\phi(0)=\phi_0\in H^1(\R^3)$.

\subsubsection{Uniqueness of solutions of GP hierarchies.}

While the existence of factorized solutions can be easily verified in the manner
outlined above, the  proof of the {\em uniqueness of solutions} of the GP hierarchy
(which encompass non-factorized solutions) is the most difficult part in 
this analysis. The proof of uniqueness of solutions to the GP hierarchy 
was originally achieved by Erd\"os, Schlein and Yau in \cite{esy1,esy2,ey} in the space 
 $\{\gamma^{(k)} \, | \, \|\gamma^{(k)}\|_{\frh^1}  \, < \, \infty\}$,
for which the authors developed highly sophisticated Feynman graph expansion methods. 
 
In \cite{klma}, Klainerman and Machedon introduced an alternative method for proving uniqueness  
in a  space of density matrices defined by the Hilbert-Schmidt type Sobolev norms 
\eqn\label{eq-gamma-norm-def-0-1}
	\| \gamma^{(k)} \|_{H^\alpha_k} \, := \,  \| S^{(k,\alpha)} \gamma^{(k)} \|_{L^2(\R^{3k} \times \R^{3k})}
	\, < \, \infty \,.
\eeqn 
While this is a different (strictly larger) space of marginal density matrices than
the one considered by Erd\"os, Schlein, and Yau, \cite{esy1,esy2},
the authors of \cite{klma} impose an additional a priori condition on  
space-time norms of the form
\eqn \label{intro-KMbound} 
     \|B_{j;k+1} \gamma^{(k+1)}\|_{L^1_t H^1_k}  \, < \, C^k \, , 
\eeqn
for some arbitrary but finite $C$ independent of $k$. 
The strategy in \cite{klma} developed to prove the uniqueness of solutions of the GP hierarchy
\eqref{eq-GP-0-0}  in $d=3$ involves the use of certain space-time bounds on density matrices 
(of generalized Strichartz type), 
and crucially employs the reformulation of a combinatorial result
in \cite{esy1,esy2} into a ``board game'' argument.
The  latter is used to organize the Duhamel expansion of solutions of the GP hierarchy
into equivalence classes of terms which leads to a significant reduction of 
the complexity of the problem. 

Subsequently, Kirkpatrick, Schlein, and Staffilani proved in \cite{kiscst} that the a priori 
spacetime bound \eqref{intro-KMbound}  
is satisfied for the cubic GP hierarchy in $d=2$, locally in time.
Their argument is based on the conservation of
energy in the original $N$-body Schr\"odinger system,
and a related a priori $H^1$-bounds for the BBGKY hierarchy in the limit $N\rightarrow\infty$
derived in \cite{esy1,esy2}, 
combined with a generalized Sobolev inequality for density matrices.

\subsection{Cauchy problem for GP hierarchies} 

In \cite{chpa2}, we began investigating the well-posedness of the
Cauchy problem for GP hierarchies, with both focusing and defocusing interactions.
We do so independently of the fact that 
it is currently not known how to rigorously derive a GP hierarchy from the
$N\rightarrow\infty$ limit of a BBGKY hierarchy with $L^2$-supercritical, attractive
interactions.
In \cite{chpa2}, we introduced the notions of {\em cubic},
{\em quintic}, {\em focusing}, or {\em defocusing GP hierarchies}, 
according to the type of NLS obtained from
factorized solutions. 

In \cite{chpa2}, we introduced the following topology on the Banach space of sequences
of $k$-particle marginal density matrices 
\eqn\label{bigG}
	\Gspace \, = \, \{ \, \Gamma \, = \, ( \, \gamma^{(k)}(x_1,\dots,x_k;x_1',\dots,x_k') \, )_{k\in\N} 
	\, | \,
	\tr \gamma^{(k)} \, < \, \infty \, \} \,.
\eeqn
Given $\xi>0$, we defined the space
\eqn
	\cH_\xi^\alpha \, = \, \{\Gamma \, | \, \| \, \Gamma \, \|_{\cH_\xi^\alpha} \, < \, \infty \, \}
\eeqn
with the norm
\eqn\label{eq-KM-aprioriassumpt-0-1}
	\| \, \Gamma \, \|_{\cH_\xi^\alpha} \, := \,
	\sum_{k\in\N} \xi^k \, \| \, \gamma^{(k)} \, \|_{H^\alpha} \,,
\eeqn
where
\eqn\label{eq-gamma-norm-def-0-0}
	\| \gamma^{(k)} \|_{H^\alpha_k} & := &  \| S^{(k,\alpha)} \gamma^{(k)} \|_{L^2} 
\eeqn  
is the norm \eqref{eq-gamma-norm-def-0-1} considered in \cite{klma}.  
If $\Gamma\in\cH_\xi^\alpha$, then $\xi^{-1}$ an upper bound on the 
typical $H^\alpha$-energy per particle;
this notion is made precise in \cite{chpa2}.
We note that small energy results are characterized by large $\xi>1$, 
while results valid without any upper
bound on the size of the energy can be proven for arbitrarily small values of $\xi>0$; 
in the latter case, one can
assume $0<\xi<1$ without any loss of generality. 
The GP hierarchy can then be written in the form
\eqn
	i\partial_t\Gamma \, + \, \opDelta_\pm\Gamma \, = \, \opB\Gamma \,,
\eeqn
with $\Gamma(0)=\Gamma_0$,
where the components of $\opDelta\Gamma$ and $\opB\Gamma$ can be read 
off from \eqref{eq-GP-0-0}. Here we have set $\kappa_0=1$.

In \cite{chpa2},  we prove the local well-posedness of solutions for 
energy subcritical focusing and defocusing  cubic 
and quintic GP hierarchies in a subspace of $\cH_\xi^\alpha$  
defined by a condition related to \eqref{intro-KMbound}.
The parameter $\alpha$ determines the regularity of the solution, 
\eqn\label{eq-alphaset-def-1}
	\alpha \, \in \, \alphaset(d,p)
	\, := \,  \left\{
	\begin{array}{cc}
	(\frac12,\infty) & {\rm if} \; d=1 \\ 
	(\frac d2-\frac{1}{2(p-1)}, \infty) & {\rm if} \; d\geq2 \; {\rm and} \; (d,p)\neq(3,2)\\
	\big[1,\infty) & {\rm if} \; (d,p)=(3,2) \,,
	\end{array}
	\right.
\eeqn     
where $p=2$ for the cubic, and $p=4$ for the quintic GP hierarchy.
Our result 
is obtained from a Picard fixed point argument, and holds for various dimensions $d$,
without any requirement on factorization.  
The parameter $\xi>0$ is determined by the initial condition, and it sets the energy scale of 
the given Cauchy problem.
In addition,  we prove lower bounds on the blowup rate for blowup solutions
of focusing GP-hierarchies in \cite{chpa2}.  
The Cauchy problem for GP hierarchies was also analyzed by the authors of
\cite{cheliu}, and the cubic GP hierarchy was derived in 
\cite{xch2} with the presence of an external trapping potential in 2D.

In the joint work \cite{chpatz1} with N. Tzirakis, 
we identify a conserved energy functional $E_1(\Gamma(t))=E_1(\Gamma_0)$ describing the
average energy per particle, 
and we prove virial identities for solutions of GP hierarchies.
In particular, we use these ingredients to prove that for $L^2$-critical
and supercritical focusing GP hierarchies, blowup occurs 
whenever $E_1(\Gamma_0)<0$ and the variance is finite.
We note that prior to  \cite{chpatz1}, no exact conserved energy functional on 
the level of the GP hierarchy was identified in any of the previous works,
including \cite{kiscst} and \cite{esy1,esy2}.

In  \cite{chpa3}, 
we  discovered an infinite family of multiplicative
energy functionals and prove that they are conserved under time evolution;
their existence is a consequence of the mean field character of GP hierarchies. 
Those conserved energy functionals allow us to  
prove global wellposedness for $H^1$ subcritical defocusing GP hierarchies,
and for $L^2$ subcritical focusing GP hierarchies. 
 
In the paper \cite{chpa4}, we prove the {\em existence} of solutions to the GP hierarchy,
without the assumption of the Klainerman-Machedon condition. This is achieved via considering 
a truncated version of the GP hierarchy (for which existence of solutions can 
be easily obtained) and showing that the limit of solutions to the truncated GP hierarchy exists as the 
truncation parameter goes to infinity, and that this limit is a solution to the GP hierarchy. Such a ``truncation-based" proof of existence of solutions  
to the GP hierarchy motivated us to try to implement a similar approach at the level of the
BBGKY hierarchy, which is what we do in this paper. 

\subsection{Main results of this paper} 
As noted above, our results in \cite{chpa2} prove the local well-posedness of solutions for spaces 
\eqn\label{eq-Gamma-solspace-1} 
	{\mathfrak W}_{\xi}^{\alpha}(I)
	\, := \,
	\{ \, \Gamma \, \in \, L^\infty_{t\in I}\cH^\alpha_\xi \; | \; \opB\Gamma
	\, \in \, L^2_{t\in I}\cH^\alpha_\xi \, \} 
	\; \; \; , \; \; \; \alpha \, \in \, \alphaset(d,p) \,,
\eeqn 
where the condition on the boundedness of the $L^2_{t\in I}\cH^\alpha_\xi$ spacetime norm 
corresponds to the condition \eqref{intro-KMbound} used by Klainerman and Machedon,
\cite{klma}.

This is a different solution space than that considered by Erd\"os, Schlein and Yau, \cite{esy1,esy2}. 
As a matter of fact, it has so far not been known if the limiting solution to the GP
hierarchy constructed by Erd\"os, Schlein, and Yau is an element of \eqref{eq-Gamma-solspace-1}
or not in dimension $d\geq3$
(for $d\leq2$, it is known to be the case, \cite{chpa,kiscst}). 
This is a central open question surrounding the well-posedness theory for 
GP hierarchies in the context of our approach developed in \cite{chpa2,chpa3,chpa4,klma}.

In this paper, we answer this question in the affirmative.
We give a
derivation of  the cubic GP hierarchy from the BBGKY hierarchy 
in dimensions $d=3$ based on the spacetime norms used in \cite{chpa2,klma}.
The main result can be formulated as follows:

%\begin{theorem} \label{thm-main-intro}
Let $d=3$, and $\delta>0$ be an arbitrary, small, fixed number. Moreover, let
\eqn
	0 \, < \, \beta \, < \, %\frac{1}{d+1+2\delta} \, = \, 
	\frac{1}{4+2\delta} \,.
\eeqn 
Suppose that 
the pair potential $\VN(x)=N^{3\beta}V(N^\beta x)$,
for $V  \in L^1(\R^3)$, is spherically symmetric,  
positive, and $\widehat V\in C^\delta(\R^3)\cap L^\infty(\R^3)$ decays rapidly outside the unit ball.

Let $(\Phi_N)_N$ denote a sequence of solutions to the
$N$-body Schr\"odinger equation  \eqref{ham1} for which we have that 
for some $0 < \xi' < 1$,  and every $N\in\N$,
$$
	\Gamma^{\Phi_N}(0) \, = \, 
	(\gamma_{\Phi_N}^{(1)}(0),\dots,\gamma_{\Phi_N}^{(N)}(0),0,0,\dots)
	\; \in \; \cH_{\xi'}^{1+\delta}
$$ 
holds at  initial time $t=0$, and moreover, that
the strong limit  
\eqn
	\Gamma_0  \, = \, \lim_{N\rightarrow\infty}\Gamma^{\Phi_N}(0)
	\, \in \, \cH_{\xi'}^{1+\delta} \,
\eeqn 
exists. 
We emphasize that $\Gamma_0$ does not need to be of factorized form.
The additonal $\delta$ amount of regularity is introduced to control the convergence
of certain terms, see section \ref{sec-bbgky-gp-1}.

We denote by
\eqn
	\Gamma^{\Phi_N}(t) \, := \, 
	(\gamma_{\Phi_N}^{(1)}(t), \dots,\gamma_{\Phi_N}^{(N)}(t), 0,0,\dots,0,\dots) \,
\eeqn
the solution to the associated BBGKY hierarchy \eqref{eq-bbgky-1} -- \eqref{eq-bbgky-2},
trivially extended by $\gamma_{\Phi_N}^{(n)}\equiv0$ for $n>N$.

%For $V$ satisfying these assumptions,
%\eqn
%	\Bra \, \Phi_N \, , \, (N+H_N)^{n} \, \Phi_N \, \Ket
%	\, \geq \, C^n \, N^{n} \, \tr(S^{(1,n)}\gamma_{\Phi_N}^{(n)})
%\eeqn
%for some positive constant $C<\infty$, and all $1\leq k\leq N$, as proven in  \cite{esy1,esy2,kiscst}.

We define the truncation operator $P_{\leq K}$ by
\eqn 
	P_{\leq K}\Gamma \, = \, (\gamma^{(1)},\dots,\gamma^{(K)},0,0,\dots) \,,
\eeqn
and let 
\eqn
	K(N) \, := \, b_0 \, \log N
\eeqn
for a suitable constant $b_0>0$.
\\

Then, the following hold for sufficiently small $0<\xi<1$:
\begin{enumerate}
\item
There exists $\Gamma \in L^\infty_{t\in [0,T]}\cH_\xi^1$ such that the limit
\eqn
	s-\lim_{N\rightarrow\infty}	
	P_{\leq K(N)}\Gamma_{\Phi_N} \, = \, \Gamma  \,
\eeqn
holds strongly in $L^\infty_{t\in [0,T]}\cH_\xi^1$.
\\

\item
Moreover, the limit
\eqn
	s-\lim_{N\rightarrow\infty}\opB_N P_{\leq K(N)}\Gamma_{\Phi_N} \, 
	= \, \opB\Gamma   
\eeqn
holds strongly in $L^2_{t\in[0,T]}\cH_\xi^1$.
\\

\item
The limit point $\Gamma \in L^\infty_{t\in [0,T]}\cH_\xi^1 $ is a mild solution to the 
cubic GP hierarchy with initial data $\Gamma_0$, satisfying
\eqn
	\Gamma(t) \, = \, U(t) \, \Gamma_0 \, + \, i \, \int_0^t U(t-s) \, \opB \, \Gamma(s) \, ds \,,
\eeqn
with $\Gamma(0)=\Gamma_0$, and $U(t):=e^{it\opDelta_\pm}$.

\end{enumerate}
%\end{theorem}

An outline of our proof is given in Section \ref{sec-mainres-1} below.
\\

\begin{remark}
We emphasize the following:  
\begin{itemize}
 
\item 
The results stated above imply that the 
$N$-BBGKY hierarchy (truncated by $P_{\leq K(N)}$ with a suitable choice of
$K(N)$) has a limit in the space introduced in \cite{chpa2}, 
which is based on the space considered
by Klainerman and Machedon in \cite{klma}.
For factorized solutions, this provides the derivation of the cubic defocusing 
NLS in those spaces.
\\

\item
In \cite{esy1,esy2,kiscst}, the limit $\gamma_{\Phi_N}^{(k)}\rightharpoonup\gamma^{(k)}$ 
of solutions to the BBGKY hierarchy
to solutions to the GP hierarchy holds in the
weak, subsequential sense, for an arbirary but fixed $k$.
In our approach, we prove strong convergence for a sequence of suitably truncated 
solutions to the BBGKY hierarchy, in an entirely different space of solutions.
An important ingredient for our construction is that   this convergence is in part
controlled by use of the parameter $\xi>0$, which is not available in  \cite{esy1,esy2,kiscst}. 
Moreover, we assume   initial data that are 
slightly more regular than of class $\cH_{\xi'}^1$.
\\

\item
We assume that the initial data has a limit, 
$\Gamma^{\phi_N}(0)\rightarrow\Gamma_0\in\cH_{\xi'}^{1+\delta}$
as $N\rightarrow\infty$,
which does not need to be factorized. We note that in \cite{esy1,esy2}, the initial data 
is assumed to be asymptotically factorized.
\\
 
\item
The method based on spacetime norms  developed in this paper 
works for the cubic case in $d=2, 3$, and is expected to have a straightforward 
generalization for the
quintic case in $d= 2$.
Our result is completely new for the cubic case in $d=3$; the other cases (of cubic and quintic GP 
in $d\leq 2$) were 
covered in \cite{chpa,kiscst}; however the mode of convergence proven here is different
and the initial data in this paper do not need to be of factorized form.
A main obstacle in treating the quintic GP hierarchy in $d=3$ is the fact 
that the currently available Strichartz estimates are not
good enough for the quintic GP hierarchy, \cite{chpa}. 
\end{itemize}
\end{remark}

\newpage

\section{Definition of the model} 
\label{sec-not}

In this section, we introduce the mathematical
model that will be studied in this paper. 
Most notations and definitions are adopted from \cite{chpa2},
and we refer to \cite{chpa2} for additional motivations and details.

\subsection{The $N$-body Schr\"odinger system}
We consider the $N$-boson Schr\"odinger equation
\eqn\label{eq-NbodySchrod-1}
	i\partial_t\Phi_N \, = \, \Big( \, - \, \sum_{j=1}^N\Delta_{x_j} \, + \, 
	\frac1 N\sum_{1\leq j<\ell \leq N} V_N(x_j-x_\ell) \, \Big) \, \Phi_N
\eeqn	
on $L^2_{Sym}(\R^{3N})$, with initial data $\Phi_N(0) \, = \, \Phi_{0,N}\in L^2_{Sym}(\R^{3N})$.
Here, $\VN(x)=N^{3\beta}V(N^\beta x)$ 
for $V  \in L^1(\R^3)$ spherically symmetric, and
positive. Moreover, we assume that $\widehat V\in C^1(\R^3)$ with rapid decay outside the unit ball.
The parameter $0<\beta<1$ is assumed to satisfy the smallness condition
\eqref{eq-betasmall-cond-0}.

Let
\eqn
	\gamma_{\Phi_N}^{(k)}
	\, := \, \tr_{k+1,\dots,N}(|\Phi_N\rangle\langle\Phi_N|) \,.
\eeqn
%and we assume $\|\gamma_{\Phi_N}^{(k)}\|_{H^{1+\delta}}<C^k$.
It is proved in  \cite{esy1,esy2,kiscst} that for $V$ satisfying the above assumptions,
\eqn\label{eq-NSchrod-aprioriEnergy-1-0}
	\Bra \, \Phi_N \, , \, (N+H_N)^{K} \, \Phi_N \, \Ket
	\, \geq \, C^K \, N^{K} \, \tr(S^{(1,K)}\gamma_{\Phi_N}^{(K)})
\eeqn
for some positive constant $C<\infty$ independent of $N,K$.
This a priori bound makes use of 
energy conservation in the $N$-body Schr\"odinger equation 
satisfied by $\Phi_N$, and will be used in the proof of our main results.   
 
\subsection{The solution spaces}
We recall the space introduced in \cite{chpa2} 
\eqn \nonumber 
	\Gspace \, := \, \bigoplus_{k=1}^\infty L^2(\R^{3k}\times\R^{3k})  
\eeqn
of sequences of marginal density matrices
\eqn \nonumber 
	\Gamma \, := \, (\, \gamma^{(k)} \, )_{k\in\N}
\eeqn
where $\gamma^{(k)}\geq0$, $\tr\gamma^{(k)} =1$,
and where every $\gamma^{(k)}(\ux_k,\ux_k')$ is symmetric in all components of $\ux_k$,
and in all components of $\ux_k'$, respectively, i.e. 
\begin{equation}\label{symmetry}
	\gamma^{(k)}(x_{\pi (1)}, ...,x_{\pi (k)};x_{\pi'( 1)}^{\prime},
	 ...,x_{\pi'(k)}^{\prime})=\gamma^{(k)}( 	x_1, ...,x_{k};x_{1}^{\prime}, ...,x_{k}^{\prime})
\end{equation}
\\
holds for all $\pi,\pi'\in S_k$. 

For brevity, we will write $\ux_k:=(x_1, \cdots, x_k)$,
and similarly, $\ux'_k:=(x'_1, \cdots, x'_k)$.
 
The $k$-particle marginals are assumed to be hermitean,
\begin{equation}\label{conj}
	\gamma^{(k)}(\ux_k;\ux_k')=\overline{\gamma^{(k)}(\ux_k';\ux_k) }.
\end{equation}
We call $\Gamma=(\gamma^{(k)})_{k\in\N}$ admissible if  
$\gamma^{(k)}=\tr_{k+1} \gamma^{(k+1)}$, that is,
\eqn 
	\gamma^{(k)}(\ux_k;\ux_k') 
	\, = \, \int dx_{k+1} \, \gamma^{(k+1)}(\ux_{k},x_{k+1};\ux_k',x_{k+1}) 
	\nonumber
\eeqn  
for all $k\in\N$.

Let $0<\xi<1$. We define
\eqn\label{eq-cHalpha-def-1} 
 	\cH_\xi^\alpha \, := \, \Big\{ \, \Gamma \, \in \, \Gspace \, \Big| \, \|\Gamma\|_{\cH_\xi^\alpha} < \, \infty \, \Big\}
\eeqn
where
\eqn \nonumber 
	\|\Gamma\|_{\cH_\xi^\alpha} \, = \, \sum_{k=1}^\infty \xi^{ k} 
	\| \,  \gamma^{(k)} \, \|_{H^\alpha_k(\R^{3k}\times\R^{3k})} \,,
\eeqn
with
\eqn\label{eq-gamma-norm-def-1}
	\| \gamma^{(k)} \|_{H^\alpha_k} & := & 
	\| S^{(k,\alpha)}  \gamma^{(k)} \|_{L^2}
\eeqn 
where $S^{(k,\alpha)}:=\prod_{j=1}^k\bra\nabla_{x_j}\ket^\alpha\bra\nabla_{x_j'}\ket^\alpha$.

\subsection{The GP hierarchy}

The main objective of the paper at hand will be to 
prove that, in the limit $N\rightarrow\infty$,
solutions of the BBGKY hierarchy converge to
solutions of an infinite hierarchy, referred to
as the Gross-Pitaevskii (GP) hierarchy.
In this section, we introduce the necessary
notations and definitions, adopting them from \cite{chpa2}. 

The cubic GP (Gross-Pitaevskii) hierarchy is given by
\eqn \label{eq-def-GP}
	i\partial_t \gamma^{(k)} \, = \, \sum_{j=1}^k [-\Delta_{x_j},\gamma^{(k)}]   
	\, + \,  \kappa_0 B_{k+1} \gamma^{(k+1)}
\eeqn
in $d$ dimensions, for $k\in\N$. Here,
\eqn \label{eq-def-b}
	B_{k+1}\gamma^{(k+1)}
	\, = \, B^+_{k+1}\gamma^{(k+1)}
        - B^-_{k+1}\gamma^{(k+1)} \, ,
\eeqn
where 
\eqn\label{eq-Bplus-GP-def-1}
	B^+_{k+1}\gamma^{(k+1)}
   = \sum_{j=1}^k B^+_{j;k+1 }\gamma^{(k+1)},
\eeqn
and 
\eqn 
	B^-_{k+1}\gamma^{(k+1)}
   = \sum_{j=1}^k B^-_{j;k+1 }\gamma^{(k+1)},
\eeqn                  
with 
\begin{align*} 
& \left(B^+_{j;k+1}\gamma^{(k+1)}\right)
(t,x_1,\dots,x_k;x_1',\dots,x_k') \\
& \quad \quad = \int dx_{k+1}  dx_{k+1}'  \\
& \quad\quad\quad\quad 
	\delta(x_j-x_{k+1})\delta(x_j-x_{k+1}' )
        \gamma^{(k+1)}(t,x_1,\dots,x_{k+1};x_1',\dots,x_{k+1}'),
\end{align*} 
and 
\begin{align*} 
& \left(B^-_{j;k+1}\gamma^{(k+1)}\right)
(t,x_1,\dots,x_k;x_1',\dots,x_k') \\
& \quad \quad = \int dx_{k+1} dx_{k+1}'  \\
& \quad\quad\quad\quad 
	  \delta(x'_j-x_{k+1})\delta(x'_j-x_{k+1}' )
        \gamma^{(k+1)}(t,x_1,\dots,x_{k+1};x_1',\dots,x_{k+1}').
\end{align*} 
 
We remark that for factorized initial data,
\eqn
	\gamma^{(k)}(0;\ux_k;\ux_k') \, = \, \prod_{j=1}^k \phi_0(x_j) \, \overline{\phi_0(x_j')} \,,
\eeqn
the corresponding solutions of the GP hierarchy remain factorized, 
\eqn
	\gamma^{(k)}(t,x_1,\dots,x_k;x_1',\dots,x_k') \,  =  \,
	\prod_{j=1}^{k} \phi(t, x_j) \, \bar{\phi}(t, x'_j) \,.
\eeqn
if the corresponding 1-particle wave function satisfies the defocusing
cubic NLS 
$$i\partial_t\phi=-\Delta\phi+\kappa_0|\phi|^2\phi \,. $$

The GP hierarchy can be rewritten in the following compact manner:
\eqn \label{chpa2-pGP}
        i\partial_t \Gamma \, + \, \opDelta_\pm \Gamma & = & \kappa_0 \, \opB \Gamma 
        \nonumber\\
        \Gamma(0) &=& \Gamma_0 \,,
\eeqn 
where
$$
	\opDelta_\pm \Gamma \, := \, ( \, \Delta^{(k)}_\pm \gamma^{(k)} \, )_{k\in\N} \, ,
        \; \; \; \; \mbox{ with }
        \Delta_{\pm}^{(k)} \, = \, \sum_{j=1}^{k} \left( \Delta_{x_j} - \Delta_{x'_j} \right)\, ,
$$
and 
\eqn \label{chpa2-B} 
	\opB \Gamma \, := \, ( \, B_{k+1} \gamma^{(k+1)} \, )_{k\in\N} \,.
\eeqn
We will also use the notation 
\begin{align*} 
	& \opB^+ \Gamma := \, ( \, B^+_{k+1} \gamma^{(k+1)} \, )_{k\in\N}, 
        \nonumber \\  
	& \opB^- \Gamma := \, ( \, B^-_{k+1} \gamma^{(k+1)} \, )_{k\in\N} \,.
\end{align*}

\subsection{The BBGKY hierarchy} 
\label{ssec-BBGKY-1}

In analogy to the compact notation for the
GP hierarchy described above,  
we introduce a
similar notation for the cubic defocusing BBGKY hierarchy.

We consider the cubic defocusing BBGKY hierarchy for the marginal density matrices,
given by
\eqn
	%\lefteqn{
	i\partial_t \gammaN^{(k)}(t) & = & \sum_{j=1}^k [-\Delta_{x_j},\gammaN^{(k)}(t)]
	+ \frac{1}{N }\sum_{1\leq j< k} [\VN(x_j-x_k  ) , \gammaN^{(k)}(t) ] 
	%}
	\nonumber\\
	&&+\frac{(N-k)}{N }
	\sum_{1\leq  j\leq k}\tr_{k+1} [\VN(x_j-x_{k+1} ) , \gammaN^{(k+1)}(t) ] \,,
	%\\ 
	\label{intro-BBGKY}
	%&&+\frac{(N-k)(N-k-1)}{N }
	%\underbrace{\tr_{k+1}\tr_{k+2}[ \VN( x_{k+1}-x_{k+2}, ) , \gammaN^{(k+2)}(t) ]}_{=0}
	%\nonumber
\eeqn
%where the term on the last line clearly vanishes, due to symmetry.  
for $k=1,\dots,n$. We extend this finite hierarchy trivially to an infinite hierarchy by
adding the terms  $\gammaN^{(k)}=0$ for $k > N$. This will allow us to treat solutions
of the BBGKY hierarchy on the same footing as solutions to the GP hierarchy.

We next introduce the following compact notation for the BBGKY hierarchy.
\eqn \label{eq-def-BBGKYN-1}
	i\partial_t \gammaN^{(k)} \, = \, \sum_{j=1}^k [-\Delta_{x_j},\gammaN^{(k)}]   
	\, + \,  \mu (B_{N} \GammaN)^{(k)}
\eeqn
for $k\in\N$. Here, we have $\gammaN^{(k)}=0$ for $k>N$, and we define
\eqn \label{eq-def-BBGKYN-Bop-1}
	(B_N\Gamma_N)^{(k)}\, := \, 
	\left\{
	\begin{array}{cc}
	B_{N;k+1}^{main} \gammaN^{(k+1)} + B_{N;k}^{error} \gammaN^{(k)} & {\rm if} \; k\leq N \\
	& \\
	0 & {\rm if} \; k>N
	\end{array}
	\right.
\eeqn
%Moreover, we set
%$B_{N;k}$ to be given by multiplication with zero for $k>N$. 
The interaction terms on the right hand side are  defined by
\eqn 
	B_{N;k+1}^{main}\gamma_N^{(k+1)}
	\, = \, B^{+,main}_{N;k+1}\gamma_N^{(k+1)}
        - B^{-,main}_{N;k+1}\gamma_N^{(k+1)} \, ,
\eeqn
and 
\eqn  
	B_{N;k}^{error}\gamma_N^{(k)}
	\, = \, B^{+,error}_{N;k}\gamma_N^{(k)}
        - B^{-,error}_{N;k}\gamma_N^{(k)} \, ,
\eeqn
where 
%\eqn 
%	B^\pm_{N;k+1}\gamma_N^{(k+1)}
%   \, = \, 
 %  B^{\pm,main}_{N;k+1}\gamma_N^{(k+1)}
%   \, + \,
%   B^{\pm,error}_{N;k+1}\gamma_N^{(k+1)}
%\eeqn
%with
\eqn
	B^{\pm,main}_{N;k+1}\gamma_N^{(k+1)}\, := \, 
	\frac{N-k}{N} \sum_{j=1}^k B^{\pm,main}_{N;j;k+1}\gamma_N^{(k+1)},
\eeqn
and 
\eqn
	B^{\pm,error}_{N;k}\gamma_N^{(k)}\, := 
	\, \frac{1}{N} \sum_{i<j}^{k} B^{\pm,error}_{N;i,j;k}\gamma_N^{(k)},
\eeqn                  
with 
\eqn
	 \lefteqn{
	 \Big( B^{+,main}_{N;j;k+1 }\gamma_N^{(k+1)}\Big)
	(t,x_1,\dots,x_k;x_1',\dots,x_k') 
	}
	\nonumber\\
	&&  
	= \int dx_{k+1}  
	 \VN(x_j-x_{k+1}) 
        \gamma_N^{(k+1)} (t,x_1,\dots,x_{k },x_{k+1};x_1',\dots,x_{k }',x_{k+1})
        \quad\quad
        \label{eq-Bmain-def-1}
\eeqn
and 
\eqn
 	\lefteqn{
	\Big(B^{+,error}_{N;i,j;k}\gamma_N^{(k)}\Big)
	(t,x_1,\dots,x_k;x_1',\dots,x_k') 
	}
	\nonumber\\
	&&
	= \, \VN(x_i-x_{j}) 
        \gamma^{(k)}(t,x_1,\dots,x_{k };x_1',\dots,x_{k }') \,.
%        \nonumber\\
%	&&
%	= \, \int dx_{k+1}  
%	 \VN(x_i-x_{j}) 
%        \gamma_N^{(k+1)}(t,x_1,\dots,x_{k },x_{k+1};x_1',\dots,x_{k }',x_{k+1}) \,,
  %      \quad\quad
        \label{eq-Berror-def-1}
\eeqn
%where the last line follows thanks to admissibility of $\gamma^{(k)}_N$. 
Moreover,
\begin{align*} 
& \left(B^{-,main}_{N;j;k+1}\gamma_N^{(k+1)}\right)
(t,x_1,\dots,x_k;x_1',\dots,x_k') \\
& \quad \quad = \int dx_{k+1} 
	 \VN(x'_j-x_{k+1}) 
        \gamma_N^{(k+1)}(t,x_1,\dots,x_{k},x_{k+1};x_1',\dots,x_{k}',x_{k+1}).
\end{align*} 
and 
\begin{align*} 
& \left(B^{-,error}_{N;i,j;k}\gamma_N^{(k)}\right)
(t,x_1,\dots,x_k;x_1',\dots,x_k') \\
& \quad \quad =  \VN(x'_i-x'_{j}) 
        \gamma^{(k)}(t,x_1,\dots,x_{k };x_1',\dots,x_{k }')\, .
%& \quad \quad = \int dx_{k+1} 
%	 \VN(x'_i-x'_{j}) 
%        \gamma_N^{(k+1)}(t,x_1,\dots,x_{k},x_{k+1};x_1',\dots,x_{k}',x_{k+1}).
\end{align*}

The advantage of this notation will be that we can treat the BBGKY hierarchy 
and the GP hierarchy on the same footing.
We remark that in all of the above definitions, we have that 
$B^{\pm,main}_{N;k}$, $B^{\pm,error}_{N;k}$, etc. are 
defined to be given by multiplication with zero for $k>N$.
 
As a consequence, we can write the 
BBGKY hierarchy compactly in the form
\eqn \label{eq-BBGKY-condensed}
        i\partial_t \GammaN \, + \, \opDelta_\pm \GammaN & = &  \opBN \GammaN 
        \nonumber\\
        \GammaN(0) &\in& \cH^\alpha_\xi  \,,
\eeqn 
where
$$
	\opDelta_\pm \GammaN \, := \, ( \, \Delta^{(k)}_\pm \gammaN^{(k)} \, )_{k\in\N} \, ,
        \; \; \; \; \mbox{ with }
        \Delta_{\pm}^{(k)} \, = \, \sum_{j=1}^{k} \left( \Delta_{x_j} - \Delta_{x'_j} \right)\, ,
$$
and 
\eqn \label{eq-BBGKY-condensed-B} 
	\opBN \GammaN \, := \, ( \, B_{N;k+1} \gammaN^{(k+1)} \, )_{k\in\N} \,.
\eeqn
In addition, we introduce the notation
\begin{align*} 
	& \opBN^+ \GammaN := \, ( \, B^+_{N;k+1} \gammaN^{(k+1)} \, )_{k\in\N}
        \nonumber \\  
	& \opBN^- \GammaN := \, ( \, B^-_{N;k+1} \gammaN^{(k+1)} \, )_{k\in\N} \,
\end{align*}
which will be convenient.

%$\;$ \\

%\newpage

\section{Statement of main results and outline of proof strategy}
\label{sec-mainres-1}

The main result proven in this paper is the following theorem.

\begin{theorem}
\label{thm-main-1}
Let $\delta>0$ be an arbitrary small, fixed number, and assume that 
\eqn\label{eq-betasmall-cond-0}
	0 \, < \, \beta \, < \, \frac{1}{4+2\delta} \,.
\eeqn 
Assume that $\Phi_N$ solves the $N$-body Schr\"{o}dinger equation \eqref{eq-NbodySchrod-1}
with initial condition $\Phi_N(t=0)=\Phi_{0,N}\in L^2(\R^{3N})$, 
where the pair potential $\VN(x)=N^{3\beta}V(N^\beta x)$,
for $V  \in  L^1(\R^3)$, is spherically symmetric,  
positive, and $\widehat V\in C^\delta(\R^3)\cap L^\infty(\R^3)$ with rapid decay outside the unit ball.

Let 
\eqn
	\Gamma^{\Phi_N} \, = \, (\gamma_{\Phi_N}^{(1)},\dots,\gamma_{\Phi_N}^{(N)},0,0,\dots)
\eeqn
denote the associated sequence of 
marginal density matrices (trivially extended by zeros), 
which solves the $N$-BBGKY hierarchy,
\eqn
	\Gamma^{\Phi_N}(t) \, = \, 
	U(t) \, \Gamma^{\Phi_N}(0) \, + \, i \, \int_0^t U(t-s) \, \opB_N\Gamma^{\Phi_N}(s) \, ds \,,
\eeqn
where $U(t):=e^{it\opDelta_\pm}$.
Furthermore, we assume that 
$\Gamma^{\Phi_{0,N}} \in\cH_{\xi'}^{1+\delta}$ for all $N$, and that
\eqn
	\Gamma_0 \, = \, \lim_{N\rightarrow\infty}\Gamma^{\Phi_{0,N}}\in\cH_{\xi'}^{1+\delta}
\eeqn
exists for some $0<\xi'<1$.

Define the truncation operator $P_{\leq K}$ by
\eqn 
	P_{\leq K}\Gamma \, = \, (\gamma^{(1)},\dots,\gamma^{(K)},0,0,\dots) \,,
\eeqn
and observe that
\eqn
	P_K\Gamma^{\Phi_N}(t) \, = \, 
	U(t) \, P_K \Gamma^{\Phi_N}(0) \, + \, i \, \int_0^t U(t-s) \, P_K\opB_N\Gamma^{\Phi_N}(s) \, ds \,.
\eeqn
Writing $\beta=:\frac{1-\delta'}4$, let
\eqn\label{eq-KnlogNchoice-0}
	K(N) \, = \, \frac{\delta'}{2\log C_0}\log N \,,
\eeqn 
and assume that 
\eqn\label{eq-xiexpcond-0}
	\xi \, < \, \min\Big\{ \, \eta^3\xi' \,,\, 
	\frac1{b_1} e^{-\frac 2{\delta'} (1-4\delta')\log C_0 } \, \Big\}
\eeqn 
where the constant
$b_1$ is as in Lemma \ref{lm-BBGKY-NSchrod-diff-2}, 
$C_0$ is as in Lemma \ref{lm-boardgame-est-1},
and $\eta$ is as in Lemma \ref{lm-BGamma-Cauchy-1}, below.  

Then, there exists  $\Gamma\in L^\infty_{t\in I}\cH_\xi^1$
with  $\opB\Gamma\in L^2_{t\in I}\cH_\xi^1$ such that the limits
\eqn
	\lim_{N\rightarrow\infty}  \| \, P_{\leq K(N)} \Gamma^{\Phi_N} 
	\, - \, %P_{\leq K(N)}
	\Gamma \, \|_{L^\infty_{t\in I}\cH_\xi^1} \, = \, 0
\eeqn
and
\eqn
	\lim_{N\rightarrow\infty} \| \,  \opB_N P_{\leq K(N)}\Gamma^{\Phi_N} 
	\, - \, %P_{\leq K(N)}
	\opB \Gamma \, \|_{L^2_{t\in I}\cH_\xi^1} \, = \, 0
\eeqn
hold, for $I=[0,T]$ with $0<T<T_0(\xi)$. 

In particular, $\Gamma$ solves the cubic GP hierarchy,
\eqn
	\Gamma(t) \, = \, U(t) \, \Gamma_0 \, + \, i \, \int_0^t U(t-s) \, \opB \, \Gamma(s) \, ds \,,
\eeqn 
with initial
data $\Gamma_0$. 
\end{theorem}

We note that the limits
$K\rightarrow\infty$ and $N\rightarrow\infty$ are taken simultaneously, and that the
smallness of the parameter $\xi>0$ is used (since small $\xi>0$ corresponds to large
energy per particle, this does not lead to any loss of generality).

In our proof, we will significantly make use of our work \cite{chpa4} which proves
the unconditional existence of solutions $\Gamma\in L^\infty_{t\in I}\cH_\xi^\alpha$ of GP hierarchies,
without  assuming $B\Gamma\in L^2_{t\in I}\cH_\xi^\alpha<\infty$.

\subsection{Outline of the proof strategy} 
 
The proof contains the following main steps:

\noindent{$\bullet$}
\underline{\bf Step 1:} 
In a first step, we construct a solution to the $N$-BBGKY hierarchy 
with truncated initial data. 

First, we recall that the $N$-BBGKY hierarchy is given by
\eqn \label{eq-outl-BBGKY}
	i\partial_t \gamma_N^{(k)} \, = \, \sum_{j=1}^k [-\Delta_{x_j},\gamma_N^{(k)}]   
	\, + \,    B_{N,k+1} \gamma_N^{(k+1)}
\eeqn
for all $k \leq N$.

Given $K$, we let $P_{\leq K}$ denote the projection operator 
\eqn
	P_{\leq K} \; : \;  \Gspace &\rightarrow&\Gspace
	\nonumber\\
	\; \Gamma_N= (\gamma_N^{(1)},\gamma_N^{(2)},\dots, \gamma_N^{(N)},0,0,\dots)
	&\mapsto&(\gamma_N^{(1)},\dots,\gamma_N^{(K)},0,0,\dots) \,,
\eeqn
and $P_{> K} = 1 - P_{\leq K}$, 
as well as $P_K:=P_{\leq K}-P_{\leq K-1}$. 

Instead of considering the solution obtained from $\Phi_N$,
we consider \eqref{eq-outl-BBGKY} with truncated initial data
$\Gamma_{0,N}^K:=P_{\leq K}\Gamma_{0,N}$, for some fixed $K$.
We will refer to solutions of this system as the $K$-truncated $N$-BBGKY hierarchy,
or $(K,N)$-BBGKY hierarchy in short. 
We note that in contrast, $\Gamma^{\Phi_N}$ solves
\eqref{eq-outl-BBGKY} with un-truncated initial data
$\Gamma_{0,N}$.

Next, we  
prove via a fixed point argument that there exists a 
unique solution of the $(K,N)$-BBGKY
hierarchy for every initial condition $\Gamma_{0,N}^K\in \cH_{\xi'}^{1+\delta}$  in the space
\eqn 
	\{ \, \Gamma_N^K \, \in \, L^\infty_{t\in I_K}\cH^{1+\delta}_{\xi'} \; | \; \opB_N\Gamma_N^K
	\, \in \, L^2_{t\in I_K}\cH^{1+\delta}_{\xi'} \, \} \,.
\eeqn 
To this end, we re-interpret $\Gamma_{0,N}^K$ as an infinite sequence, extended by
zeros for elements $(\Gamma_{0,N}^K)^{(k)}=0$ with\footnote{
We observe that then, \eqref{eq-outl-BBGKY} determines a closed, infinite sub-hierarchy, 
for initial data $\gamma_N^{(k)}(0) =  0 $,  for $k> K$, which has the trivial solution
\eqn 
	(\gamma_N^K)^{(k)}(t) \, = \, 0 
	\; \; \; \; ,  \; \; \; t\in I=[0,T] \; \; \; , \; \; \; k>K \,.
\eeqn
}
$k>K$.
%We observe that then, \eqref{eq-outl-BBGKY} determines a closed, infinite sub-hierarchy, 
%for initial data $\gamma_N^{(k)}(0) =  0 $,  for $k> K$, which has the trivial solution
%\eqn 
%	(\gamma_N^K)^{(k)}(t) \, = \, 0 
%	\; \; \; \; ,  \; \; \; t\in I=[0,T] \; \; \; , \; \; \; k>K \,.
%\eeqn
To obtain this result, we need to require that given $K$, $N$ is large enough that the condition
\eqn\label{eq-KlogNbound-0-0-1}
	K \, < \, \frac{\delta'}{\log C_0}\log N
\eeqn
is satisfied. Clearly, the choice \eqref{eq-KnlogNchoice-0} complies with this condition.
The condition \eqref{eq-KlogNbound-0-0-1} is needed by the Lemma \ref{lm-BGamma-Cauchy-1}
proven in the Appendix, which is crucial for the fixed point argument used in this part of the
proof. It uses the Klainerman-Machedon boardgame argument to account for 
the part $B^{main}_N$ in the interaction operator $B_N$; to also accommodate the part $B_N^{error}$,
the condition \eqref{eq-KlogNbound-0-0-1} is used.

Hence, we have obtained solutions $\Gamma_N^K(t)$ of the  BBGKY hierarchy,
\eqn \label{eq-outl-BBGKY-special} 
	i\partial_t\Gamma_N^K \, = \, \opDelta_\pm\Gamma_N^K \, + \,   \opB_N\Gamma_N^K \,,
\eeqn 
for the truncated initial data
\eqn 
	\Gamma_N^K(0) \, = \, P_{\leq K}\Gamma_N(0) 
	\, = \, (\gamma_N^{(1)}(0),\dots,\gamma_N^{(K)}(0),0,0,\dots)
\eeqn
for an arbitrary, large, fixed $K \leq N$, and where  
component $(\Gamma_N^K)^{(m)}(t) = 0$ for the $m$-th component, for
all $m > K$. 
By the Duhamel formula, the solution of   \eqref{eq-outl-BBGKY-special}  is given by
\eqn \label{eq-outl-Duhamel-GammaK}
	\Gamma_N^K(t) \, = \, U(t)\Gamma_N^K(0) \, + \, i \,  \int_0^t U(t-s) \, \opB_N\Gamma_N^K(s) \, ds
\eeqn
for initial data $\Gamma_N^K(0)=P_{\leq K}\Gamma_N(0)$.  

%For a fixed scale $0 < \xi < 1$, 
%it is sufficient to iterate the Duhamel formula   \eqref{eq-outl-Duhamel-GammaK}
%for $\Gamma_N^K$ only finitely many times, 
%in order to obtain a fully explicit solution to \eqref{eq-outl-BBGKY-special} for fixed $K$
%that satisfies
%\eqn\label{eq-outl-BGamma-Duhj-combin}  
%	\| \,  \Gamma_N^K
%	\, \|_{L^\infty_{t\in I}\cH^{1+\delta}_{\xi} }   
%	\;\; , \;\;\;
%	\| \, \opB_N \Gamma_N^K
%	\, \|_{L^2_{t\in I}\cH^{1+\delta}_{\xi} }  
%	\, \leq \, C(T,\xi) \, \|\Gamma_0\|_{\cH^{1+\delta}_{\xi}}\,.
%\eeqn    

\noindent{$\bullet$}
\underline{\bf Step 2:} 
In this step, we take the limit $N \rightarrow \infty$ of the solution $\Gamma_N^{K(N)}$ 
to \eqref{eq-outl-BBGKY-special} which was obtained in Step 1, for some 
sequence $K(N)\rightarrow\infty$ as $N\rightarrow\infty$ that could be quite arbitrary.
However, to comply with other parts of the proof, the choice  
\eqref{eq-KnlogNchoice-0} is made for $K(N)$.

To this end, we invoke the solution $\Gamma^K$ of the GP hierarchy with 
truncated initial data, $\Gamma^K(t=0)=P_{\leq K}\Gamma_0\in\cH_\xi^1$.
In \cite{chpa4}, we proved the existence of a solution  $\Gamma^K$
that satisfies the $K$-truncated GP-hierarchy in  integral form, 
\eqn \label{eq-outl-GP-Duhamel-special-2}
	\Gamma^K(t) \, = \, U(t)\Gamma^K(0) \, + \, i \,  \int_0^t U(t-s) \, \opB\Gamma^K(s) \, ds
\eeqn 
where $(\Gamma^K)^{(k)}(t) = 0$ for all $k > K$. Moreover, 
it is shown in \cite{chpa4} that this solution satisfies
$\opB\Gamma^K\in L^2_{t\in I}\cH^{1}_{\xi}$.

We then prove the following convergence:
\begin{enumerate} 
\item[(a)] In the limit $N\rightarrow\infty$, $ \Gamma_N^{K(N)} $ satisfies
\eqn 
	\lim_{N\rightarrow \infty} \| \Gamma_N^{K(N)} 
	\, - \, \Gamma^{K(N)} \|_{ L^\infty_t \cH^{1}_{\xi}}
	\, = \, 0 \,.  
\eeqn

\item[(b)] In the limit  $N\rightarrow\infty$, $\opB_N \Gamma_N^{K(N)}$ satisfies
\eqn 
	\lim_{N \rightarrow \infty} \| {\opB}_N \Gamma_N^{K(N)} -
	\opB\Gamma^{K(N)}\|_{L^2_t \cH^{1}_{\xi}} \, = \, 0 \,.
\eeqn 
\end{enumerate}
The proof of these limits makes use of the $\delta$ amount of extra regularity
of the initial data $\Gamma_0,\Gamma_{0,N}\in\cH_{\xi'}^{1+\delta}$ beyond $\cH_{\xi'}^{1}$.

%Hence we can summarize this step in the following way: 
%as $N \rightarrow \infty$, $\Gamma_N^K \rightarrow \Gamma^K \in \L^2_t \cH^{\alpha}_{\xi}$ 
%such that  
%\eqn \label{eq-outl-GP-Duhamel-special-2}
%	\Gamma^K(t) \, = \, U(t)\Gamma^K(0) \, + \, i \,  \int_0^t U(t-s) \, \opB\Gamma^K(s) \, ds
%\eeqn 
%with $\Gamma^K(0) = P_{\leq K} \Gamma_0 \in \cH^{\alpha}_{\xi} $,
%where 
%$$
%	(\Gamma^K)^{(k)}(t) = 0 \mbox{ for all } k > K
%$$ 
%and 
%$$
%	\opB \Gamma^K \in L^2_t \cH^{\alpha}_{\xi}.
%$$

\noindent{$\bullet$}
\underline{\bf Step 3:} 
We compare the solution $\Gamma_N^K$ of the $K$-truncated $N$-BBGKY hierarchy to 
the the truncated solution $P_{\leq K(N)}\Gamma^{\Phi_N}$ of the $N$-BBGKY hierarchy.
Notably, both have the same value at $t=0$, given by $P_{\leq K(N)}\Gamma_{0,N}$.
We prove that 
\eqn
	\lim_{N\rightarrow\infty}
	\| \Gamma_N^{K(N)} \, - \, P_{\leq K(N)}\Gamma^{\Phi_N} \|_{L^\infty_{t\in I}
	\cH_\xi^1} \, = \, 0 \,.
\eeqn
and
\eqn
	\lim_{N\rightarrow\infty}
	\|B_{N} \Gamma_N^{K(N)} \, - \, B_N P_{\leq K(N)}\Gamma^{\Phi_N} \|_{L^2_{t\in I}
	\cH_\xi^1} \, = \, 0 \,.
\eeqn
The proof of this limit involves the a priori energy bounds for the $N$-body Schr\"{o}dinger system
\eqref{eq-NSchrod-aprioriEnergy-1-0} established in \cite{esy1,esy2,kiscst}.
%Accordingly, in this step, the regularity $\alpha=1$ is essential.

The norm differences considered here can be bounded by $O(K^{a_1}\xi^K N^{a_2})$  at finite $N$, 
for 
some positive constants $a_1,a_2$. We may choose 
$K(N)= b\log N$, with the constant $b$ small enough to satisfy
\eqref{eq-KlogNbound-0-0-1}; this is accomplished by \eqref{eq-KnlogNchoice-0}.
We then make use of the freedom to choose the parameter $\xi>0$ to be sufficiently small;
the condition \eqref{eq-xiexpcond-0} suffices to obtain $O((K(N))^{a_1}\xi^{K(N)} N^{a_2})\rightarrow0$ as $N\rightarrow\infty$.
\\

\noindent{$\bullet$}
\underline{\bf Step 4:} 
Finally, we determine the limit $N \rightarrow \infty$ of  $\Gamma^{K(N)}$ 
from Step 2, obtaining that: 
\begin{enumerate} 
\item[(i)] The strong limit $\lim_{N \rightarrow \infty} \Gamma^{K(N)}$  exists 
in $L^\infty_t \cH^{1}_{\xi}$, and satisfies
\eqn 
	\lim_{N\rightarrow \infty} \Gamma^{K(N)} \, =
	\, \Gamma \in L^\infty_t \cH^{1}_{\xi},  
\eeqn
where $\Gamma$ is a solution to the full GP hierarchy \eqref{eq-def-GP}
with initial data $\Gamma_0$.
\\

\item[(ii)] 
In addition, the strong limit $\lim_{N \rightarrow \infty} {\opB} \Gamma^{K(N)}$ exists
in $L^2_t \cH^{1}_{\xi}$, and satisfies
\eqn 
	\lim_{N \rightarrow \infty} {\opB} \Gamma^{K(N)} 
	\, = \, \opB \Gamma \;\;\; \in \, L^2_t \cH^{1}_{\xi}.
\eeqn 
\end{enumerate}
The results of Step 4 were  proven in
our earlier work \cite{chpa4}.
\\

\newpage

\section{Local well-posedness for the $(K,N)$-BBGKY hierarchy}

In this section, we prove the local well-posedness of the Cauchy problem for the
$K$-truncated $N$-BBGKY hierarchy, which we refer to as the $(K,N)$-BBGKY hierarchy
for brevity. In the sequel, we will have $d=2,3$.

\begin{lemma}
\label{lm-KNBBGKY-wp-1}
Assume that $N$ is sufficiently large, and in particular,
given $K\in\N$ and $\beta=:\frac{1-\delta'}4$, that
\eqn\label{eq-KnlogNchoice-0-1}
	K \, < \, \frac{\delta'}{\log C_0}\log N \,
\eeqn 
holds.
Assume that $\Gamma_{0,N}^K =P_{\leq K}\Gamma_{0,N}\in \cH_{\xi'}^{1+\delta}$ for 
some $0 < \xi' < 1$ and $\delta\geq0$. 
%for $\alpha\in\alphaset(d)$.
Then, there exists a unique solution 
$\Gamma_N^K\in L^\infty_{t\in I}\cH_\xi^{1+\delta}$ of \eqref{eq-outl-Duhamel-GammaK}
for $I=[0,T]$ with $T>0$ sufficiently small, and independent of $K,N$.
In particular,
$\opB_N\Gamma_N^K\in L^2_{t\in I}\cH_\xi^{1+\delta}$.
Moreover,
\eqn
	\|\Gamma_N^K\|_{L^\infty_{t\in I}\cH_\xi^{1+\delta}}
	\, \leq \,
	 C_2(T,\xi,\xi') \, \|\Gamma_{0,N}^K\|_{\cH_{\xi'}^{1+\delta}}
\eeqn
and
\eqn\label{eq-ThetaNK-L2Hxi-bound-0-1}
	\|\opB_N\Gamma_N^K\|_{L^2_{t\in I}\cH_\xi^{1+\delta}} 
 	\, \leq \, C_2(T,\xi,\xi') \, \|\Gamma_{0,N}^K\|_{\cH_{\xi'}^{1+\delta}}
\eeqn
hold for $0 < \xi < \xi'$ sufficiently small (it is sufficient that $0 < \xi < \eta \xi'$ with 
$\eta$ specified in Lemma \ref{lm-BGamma-Cauchy-1} below). 
The constant $C_2=C_2(T,\xi,\xi')$ is
independent of $K,N$.

Furthermore, $(\Gamma_N^K(t))^{(k)}=0$ for all $K< k\leq N$, and
all $t\in I$.
\end{lemma}

\prf
To obtain local well-posedness of the Cauchy problem for the $(K,N)$-BBGKY hierarchy,
we consider the map
\eqn
	\cM_N^K(\widetilde\Theta^{K-1}) \, := \,   \opB_NU(t)\Gamma_{N,0}^{K} \, + \, i 
	\int_0^t \opB_N U(t-s)  \widetilde\Theta^{K-1}(s) \,,
\eeqn
where $P_{\leq K-1}\widetilde\Theta^{K-1}=\widetilde \Theta^{K-1}$ on the subspace
${\rm Ran}(P_{\leq K})\cap L^2_{t\in I}\cH_\xi^{1+\delta}\subset L^2_{t\in I}\cH_\xi^{1+\delta}$.
Using the $K$-truncated Strichartz estimate in  Proposition \ref{prp-KStrichartz-1},
we find that 
\eqn
	\lefteqn
	{
	\|\cM_N^K(\widetilde\Theta^{K-1}_1)
	-\cM_N^K(\widetilde\Theta^{K-1}_2)\|_{L^2_{t\in I}\cH_\xi^{1+\delta}}
	}
	\nonumber\\
	&\leq&
	\Big\|\int_0^t ds \Big\|
	\opB_N U(t-s) (  \widetilde\Theta_1^{K-1}
	-\widetilde\Theta_2^{K-1})(s) \, \Big\|_{\cH_\xi^{1+\delta}}
	\, \Big\|_{L^2_{t\in I}}
	\nonumber\\
	&\leq&
	\int_0^T ds \Big\|
	\opB_N U(t-s) (  \widetilde\Theta_1^{K-1}-
	\widetilde\Theta_2^{K-1})(s)\Big\|_{L^2_{t\in I}\cH_\xi^{1+\delta}}
	\nonumber\\
	&\leq&
	C_2(K) \, \xi^{-1}
	\int_0^T ds \Big\| ( \widetilde\Theta_1^{K-1}
	-\widetilde\Theta_2^{K-1})(s) \Big\|_{\cH_\xi^{1+\delta}}
	\nonumber\\
	&\leq&
	C_2(K) \, \xi^{-1} \, T^{\frac12} \,
	\Big\| \widetilde\Theta_1^{K-1}
	-\widetilde\Theta_2^{K-1} \Big\|_{L^2_{t\in I}\cH_\xi^{1+\delta}} \,.
\eeqn
Thus, for $(T(K))^{\frac12}<\frac{\xi}{2C_2(K)}$, we find that
$\cM_{N}^K$ is a contraction on $L^2_{t\in I}\cH_\xi^{1+\delta}$. 
By the fixed point principle,
we obtain a unique solution $\Theta_N^{K-1} \in L^2_{t\in I}\cH_\xi^{1+\delta}$
with $\Theta_N^{K-1}=P_{\leq K-1}\Theta_N^{K-1}$ satisfying
\eqn
	\Theta_N^{K-1}(t) \, = \, \opB_N U(t) \Gamma_{N,0}^{K} \, + \, i 
	\int_0^t \opB_N U(t-s)  \Theta_N^{K-1}(s) \, ds \,.
\eeqn
In particular,
\eqn
	\|\Theta_N^{K-1}\|_{L^2_{t\in I}\cH_\xi^{1+\delta}} 
	\, \leq \, \|\opB_NU(t)\Gamma_{N,0}^K\|_{L^2_{t\in I}\cH_\xi^{1+\delta}} 
	\, + \, C_2(K) \, \xi^{-1} \,  T^{\frac12} \|\Theta_N^{K-1}\|_{L^2_{t\in I}\cH_\xi^{1+\delta}} 
	\; \;
\eeqn
and use of Proposition \ref{prp-KStrichartz-1} implies that
\eqn\label{eq-ThetaNK-L2Hxi-bound-1}
	\|\Theta_N^{K-1}\|_{L^2_{t\in I}\cH_\xi^{1+\delta}} 
 	\, \leq \, \frac{C_2(K) \, \xi^{-1} \,  }{1-C_2(K)\, \xi^{-1} \, T^{\frac12}} 
	\, \|\Gamma_{0,N}^{K}\|_{\cH_\xi^{1+\delta}}
\eeqn
holds.

Next, we let
\eqn
	\Gamma_N^K(t)  \, := \, U(t)\Gamma_{N,0}^K \, + \, i 
	\int_0^t U(t-s)  \Theta_N^{K-1}(s) \, ds \,.
\eeqn
Clearly,
\eqn\label{eq-ThetaNK-L2Hxi-bound-0-1-0}
	\|\Gamma_N^K\|_{L^\infty_{t\in I}\cH_\xi^{1+\delta}}
	&\leq& \|\Gamma_{0,N}^K\|_{L^\infty_{t\in I}\cH_\xi^{1+\delta}}
	\, + \, T^{\frac12} \, \|\Theta^{K-1}_N \|_{L^2_{t\in I}\cH_\xi^{1+\delta}} 
	\nonumber\\
	&\leq&
	 \frac{1}{1-C_2(K) \, \xi^{-1} \,  T^{\frac12}} \, \|\Gamma_{0,N}^K\|_{\cH_\xi^{1+\delta}}
\eeqn
from \eqref{eq-ThetaNK-L2Hxi-bound-1}.
Comparing the right hand sides of $B_N\Gamma_N^K$ and $\Theta_N^{K-1}$, we conclude 
that 
\eqn
	B_N\Gamma_N^K \, = \, \Theta_N^{K-1} \,
\eeqn
holds, and that 
\eqn\label{eq-GammaNK-Duh-def-1}
	\Gamma_N^K(t)  \, = \, U(t)\Gamma_{N,0}^K \, + \, i 
	\int_0^t U(t-s)  \opB_N\Gamma_N^K(s) \, ds
\eeqn
is satisfied, with $B_N\Gamma_N^K \in L^2_{t\in I}\cH_\xi^{1+\delta}$.
So far, we have established well-posedness of solutions of the $(K,N)$-BBGKY hierarchy
for $t\in[0,T]$ with $T<T_0(K,\xi)$. We can piece those solutions together,
in order to extend them
to longer time intervals.
 
In particular, we can prove that \eqref{eq-ThetaNK-L2Hxi-bound-0-1-0} can be enhanced
to an estimate with both $C_2$ and $T_0$ independent of $K$, provided $N$ is large 
enough for \eqref{eq-KnlogNchoice-0-1} to hold.
In this case, we observe that applying $\opB_N$ to \eqref{eq-GammaNK-Duh-def-1},
we find
\eqn\label{eq-BNGammaNK-Duh-def-1}
	\opB_N\Gamma_N^K(t)  \, = \, \opB_N U(t)\Gamma_{N,0}^K \, + \, i 
	\int_0^t \opB_N U(t-s)  \opB_N\Gamma_N^K(s) \, ds \,.
\eeqn
It is easy to verify that the assumptions of Lemma \ref{lm-BGamma-Cauchy-1}
in the Appendix are
satisfied for
\eqn
	\widetilde\Theta_N^K \, := \, \opB_N \, \Gamma_N^K 
	\; \; \; \; , \; \; \; \;
	\Xi_N^K \, := \, \opB_N U(t) \Gamma_{0,N}^K \,.
\eeqn 
We assume that  
\eqn\label{eq-xiparm-def-1}
	\xi \, < \, \eta \, \xi'' \, < \, \eta^2 \xi' 
\eeqn
where $0<\eta<1$ is as in  Lemma \ref{lm-BGamma-Cauchy-1}.
Then, Lemma  \ref{lm-BGamma-Cauchy-1} implies that
\eqn\label{eq-ThetaNK-L2Hxi-bound-0-1-1}
	\|\opB_N\Gamma_N^K\|_{L^2_{t\in I}\cH_\xi^{1+\delta}} 
 	& \leq &
	C(T,\xi,\eta) \, \|\opB_N U(t) \Gamma_{0,N}^K\|_{L^2_{t\in I}\cH_{\xi''}^{1+\delta}}
	\nonumber\\
	&\leq& C_2(T,\xi,\eta) \, \|\Gamma_{0,N}^K\|_{\cH_{\xi'}^{1+\delta}}
\eeqn
holds for a constant $C_2=C_2(T,\xi,\eta)$  
independent of $K,N$, and for $T<T_0(\xi,\eta)$, if $N$ is sufficiently large. 

It remains to prove that $(\Gamma_N^K(t))^{(k)}=0$ for all $K< k\leq N$, and
all $t\in I$. To this end, we first note that
\eqn \label{eq-BPprop}
	(\opB_N P_{\leq K} \, - \, P_{\leq K-1} \opB_N) \,  \Gamma^N_K
	\, = \, 0 \,,
\eeqn
as one easily verifies based on the componentwise definition of $\opB_N$
in \eqref{eq-Bmain-def-1} and \eqref{eq-Berror-def-1}. Hence, in particular, 
$$ (P_{>K} \opB_N - \opB_N P_{>K+1}) \, \Gamma^N_K = 0,$$ 
thanks to which we observe that $P_{>K}\Gamma_N^K$ by itself satisfies
a closed  sub-hierarchy of the $N$-BBGKY hierarchy,
\eqn	
	i \partial_t (P_{>K}\Gamma_N^K)
	\, = \, \opDelta_\pm (P_{>K}\Gamma_N^K) \, + \, \opB_N (P_{>K+1}\Gamma_N^K) \,,
\eeqn
where clearly,
\eqn
	P_{>K+1}\Gamma_N^K = P_{>K+1}(P_{>K}\Gamma_N^K) \,,
\eeqn
with initial data
\eqn
	(P_{>K}\Gamma_N^K)(0) \, = \, P_{>K}(\Gamma_N^K(0)) \, = \, 0 \,.
\eeqn
Here we recall that the initial data is truncated for $k> K$.

Accordingly, by the same argument as above, there exists a unique solution
$(P_{>K}\Gamma_N^K)\in L^\infty_{t\in I}\cH_\xi$ 
with $\opB_N(P_{>K+1}\Gamma_N^K)\in L^2_{t\in I}\cH_\xi$ such that
\eqn
	\|\opB_N(P_{>K+1}\Gamma_N^K)\|_{ L^2_{t\in I}\cH_\xi} \, \leq \, 
	C_2(T,\xi,\eta) \, \|(P_{>K}\Gamma_N^K)(0)\|_{\cH^{1+\delta}_{\xi'}}
	\, = \, 0 \,,
\eeqn
for $\xi<\eta^2\xi'$. Moreover,
\eqn
	\|P_{>K}\Gamma_N^K\|_{ L^\infty_{t\in I}\cH_\xi} \, \leq \, 
	C_1(T,\xi,\eta) \, \|(P_{>K}\Gamma_N^K)(0)\|_{\cH^{1+\delta}_{\xi'}}
	\, = \, 0 \,.
\eeqn
This implies that $(P_{>K}\Gamma_N^K)(t)=0$ for $t\in I$, as claimed.
\endprf

$\;$ \\

%\newpage

\section{From $(K,N)$-BBGKY to  $K$-truncated GP hierarchy}
\label{sec-bbgky-gp-1}

In this section, we control the limit $N\rightarrow\infty$ of the truncated BBGKY
hierarchy, at fixed $K$. 

\begin{proposition}\label{prp-GammaKN-conv-1} 
Assume that $V_N(x)=N^{3\beta}V(N^\beta x)$ with $\widehat V\in C^\delta\cap L^\infty$
for some arbitrary but fixed, small $\delta>0$.
Moreover, assume that $\Gamma^K\in {\mathfrak W}_{\xi}^{1+\delta}(I)$ 
(see \eqref{eq-Gamma-solspace-1})
is the solution of the GP hierarchy with truncated initial data
$\Gamma_0^K=P_{\leq K}\Gamma_0\in\cH_\xi^{1+\delta}$ constructed in \cite{chpa4}.

Let $\Gamma_N^K$ solve the $(K,N)$-BBGKY hierarchy with initial data
$\Gamma_{0,N}^K:= P_{\leq K}\Gamma_{0,N}\in\cH_{\xi'}^{1+\delta}$.
Let, for $\beta=\frac{1-\delta'}{4}$,
\eqn\label{eq-KNdef-5-1}
	K(N) \, := \, \frac{\delta'}{2\log C_0} \, \log N
\eeqn
so that \eqref{eq-KnlogNchoice-0-1} is satisfied.
Then, as $N\rightarrow\infty$,
the strong limits
\eqn\label{eq-GammaK-Nlim-1}
	\lim_{N\rightarrow\infty}\| \, \Gamma^{K(N)}_N 
	\, - \, \Gamma^{K(N)} \, \|_{L^\infty_{t\in[0,T]}\cH_\xi^1} \, = \, 0
\eeqn
and
\eqn\label{eq-BGammaK-Nlim-1}
	 \lim_{N\rightarrow\infty} \| \, \opB_{N}\Gamma^{K(N)}_N 
	 \,  - \, \opB\Gamma^{K(N)} \, \|_{L^2_{t\in[0,T]}\cH_\xi^1}
	 \, = \, 0
\eeqn
hold, for $0<T<T_0(\xi)$. 

\end{proposition}

\prf
In \cite{chpa4}, we constructed a solution $\Gamma^K$ of the full GP hierarchy
with truncated initial data, $\Gamma(0)=\Gamma_0^K\in\cH_\xi^{1+\delta}$, 
satisfying the following:
For an arbitrary fixed $K$,  $\Gamma^K$ satisfies the GP-hierarchy  in  integral 
representation, 
\eqn \label{eq-outl-GP-Duhamel-special-1}
	\Gamma^K(t) \, = \, U(t)\Gamma^K_0 \, + \, i \,  \int_0^t U(t-s) \, \opB\Gamma^K(s) \, ds \,,
\eeqn 
and in particular, $(\Gamma^K)^{(k)}(t) = 0$ for all $k > K$. 

Accordingly, we have
\eqn
	\lefteqn{
	\opB_N\Gamma_N^K-\opB\Gamma^K
	}
	\nonumber\\
	&=&
	\opB_N U(t)\Gamma_{0,N}^K-\opB U(t) \Gamma_0^K 
	\nonumber\\
	&&
	+\, i   \int_0^t \, \big( \, \opB_N U(t-s) \opB_N\Gamma_N^K 
	\, - \, \opB U(t-s) \opB \Gamma^K  \, \big)(s) ds \, 
	\nonumber\\
	&=&
	(\opB_N \, - \, \opB) U(t) \Gamma_{0,N}^K 
	\, + \,  \opB U(t)(\Gamma_{0,N}^K- \Gamma_0^K) 
	\nonumber\\
	&&
	+\,  i  \, \int_0^t \big( \, \opB_N -\opB \big) U(t-s) \opB\Gamma^K (s) \, ds 
	\nonumber\\
	&&
	+\, i  \int_0^t \opB_N  U(t-s) \big( \opB_N\Gamma_N^K \, - \,  
	\opB\Gamma^K  \, \big)(s) \, ds \,.
\eeqn
Here, we observe that for $N$ sufficiently large, 
we can apply Lemma  \ref{lm-BGamma-Cauchy-1} with
\eqn
	\widetilde\Theta_N^K &:=& \opB_N\Gamma_N^K-\opB\Gamma^K
\eeqn
and
\eqn
	\Xi_N^K &:=& (\opB_N \, - \, \opB) U(t) \Gamma_{0,N}^K 
	\, + \,  \opB U(t)(\Gamma_{0,N}^K- \Gamma_0^K) 
	\nonumber\\
	&&
	+\,  i  \, \int_0^t \big( \, \opB_N -\opB \big) U(t-s) \opB\Gamma^K (s) \, ds \,.
\eeqn
Given $\xi'$, we introduce  parameters $\xi,\xi'',\xi'''$ satisfying
\eqn\label{eq-xiparm-def-1}
	\xi \, < \, \eta \, \xi'' \, <  \, \eta^2 \, \xi''' \, <  \, \eta^3 \xi' 
\eeqn
where $0<\eta<1$ is as in  Lemma \ref{lm-BGamma-Cauchy-1}.
Accordingly,  Lemma \ref{lm-BGamma-Cauchy-1} implies that
\eqn
	\lefteqn{
	\|\opB_N\Gamma_N^K-\opB\Gamma\|_{L^2_{t\in I}\cH^1_\xi}
	}
	\nonumber\\ 
	&\leq& 
	C_2(T,\xi,\xi'') 
	\Big( \,  \|B U(t)(\Gamma_{0,N}^K- \Gamma_0^K)\|_{L^2_{t\in I}\cH^1_{\xi''}}
	\, + \, R^K(N) \, \Big)
	\nonumber\\ 
	&\leq& 
	C_1(T,\xi,\xi',\xi'') 
	\Big( \, \| \Gamma_{0,N}^K- \Gamma_0^K\|_{L^2_{t\in I}\cH^1_{\xi'}}
	\, + \, R^K(N) \, \Big)  \,,
\eeqn
where we used Lemma \ref{lm-global-free-Strichartz-1} to pass to the last line. Here,
\eqn
	R^K(N) \, = \, R_1^K(N) \, + \, R_2^K(N) \,,
\eeqn
with
\eqn
	R_1^K(N) \, := \,
	\|(\opB_N \, - \, \opB) U(t) \Gamma_{0,N}^K\|_{L^2_{t\in I}\cH^1_{\xi''}}
\eeqn
and
\eqn
	R_2^K(N) \, := \,  
	\Big\| \, \int_0^t \big( \, \opB_N -\opB \big) U(t-s) \opB\Gamma^K (s) \, ds
	\, \Big\|_{L^2_{t\in I}\cH^1_{\xi''}} \,.
\eeqn
Next, we consider the limit $N\rightarrow\infty$ with $K(N)$ as given in \eqref{eq-KNdef-5-1}.
We choose $K(N)$ in this manner for it to be compatible with  
\eqref{eq-KnlogNchoice-0-1}, which is necessary for results in other sections.

To begin with, we note that
\eqn
	\lim_{N\rightarrow \infty}
	\| \Gamma_{0,N} - \Gamma_0\|_{\cH^{1+\delta}_{\xi'}}
	\, = \, 0 \,.
\eeqn
Including the truncation at $K(N)$, it is easy to see that  
\eqn
	%\lefteqn{
	\lim_{N\rightarrow \infty}
	\| \Gamma_{0,N}^{K(N)}- \Gamma_0^{K(N)}\|_{\cH^{1+\delta}_{\xi'}}
	%}
	%\nonumber\\
	&=&
	\lim_{N\rightarrow \infty}
	\| \, P_{\leq K(N)} \, ( \, \Gamma_{0,N}- \Gamma_0 \, ) \, \|_{\cH^{1+\delta}_{\xi'}}
	\nonumber\\
	&\leq&
	\lim_{N\rightarrow \infty}
	\| \,   \Gamma_{0,N}- \Gamma_0 \, \|_{\cH^{1+\delta}_{\xi'}}
	\nonumber\\
	&=&
	0 \,
\eeqn
follows.

To control $R^{K(N)}(N)$, we  invoke Lemma \ref{lm-R1bound-1} below, which implies
that for an arbitrary but fixed $\delta>0$,
\eqn\label{eq-tildeR1-def-1}
	\lim_{N\rightarrow\infty} R_1^{K(N)}(N)
	& \leq &\lim_{N\rightarrow\infty}
	C_{V,\delta} \, \xi^{-1} \,  N^{-\delta\beta} \, 
	\| \Gamma_{0,N}^{K(N)}\|_{L^2_{t\in I}\cH^{1+\delta}_{\xi'}}
	\nonumber\\
	%&\leq&
	%(\lim_{N\rightarrow\infty} N)
	%C_0\, \xi^{-1} \, 
	%(\lim_{N\rightarrow\infty}
	 %\|\Gamma_{0,N}^{K(N)}\|_{\cH^1_{\xi'}})
	%\nonumber\\
	&=&  0\,,  
\eeqn
for a constant $C_{V,\delta}$ that depends only on $V$ and $\delta$,
since 
\eqn
	\lim_{N\rightarrow\infty}\|\Gamma_{0,N}^{K(N)}-\Gamma_0\|_{\cH^{1+\delta}_{\xi'}} \, = \, 0\,,
\eeqn
and $\|\Gamma_0\|_{\cH^{1+\delta}_{\xi'}}<\infty$.

Moreover, invoking Lemma \ref{lm-R1bound-2} below, we find 
\eqn\label{eq-tildeR2-def-1}
	\lim_{N\rightarrow\infty}  R_2^{K(N)}(N)
	& \leq &
	\lim_{N\rightarrow\infty}
	C_{V,\delta} \, \xi^{-1} \,  N^{-\delta\beta} \,  
	\|\opB\Gamma^{K(N)}   \|_{L^2_{t\in I}\cH_{\xi'''}^{1+\delta}} 
	\nonumber\\
	&=&0 \,, 
\eeqn
because
\eqn
	\|\opB\Gamma^{K(N)}\|_{L^2_{t \in I}\cH^{1+\delta}_{\xi'''}} \, < \, 
	C(T,\xi''',\xi') \, \|\Gamma_{0}\|_{\cH^{1+\delta}_{\xi'}}
\eeqn
is uniformly bounded in $N$, as shown in \cite{chpa4}. 
\endprf

\begin{lemma}
\label{lm-R1bound-1}
Let $\delta>0$ be an arbitrary, but fixed, small number.
Assume that  
$V_N(x)=N^{3\beta}V(N^\beta x)$ with $\widehat V\in C^\delta\cap L^\infty$. 
Then, with $\xi<\eta \, \xi''$ as in \eqref{eq-xiparm-def-1},
\eqn
	%\lefteqn{
	\|(\opB_N \, - \, \opB) U(t) \Gamma_{0,N}^K\|_{L^2_{t\in \R}\cH^1_\xi}
	%}
	%\nonumber\\
	%&&
	\, < \, C_{V,\delta} \,  \xi^{-1} \, N^{-\delta\beta} \, 
	\|\Gamma_{0,N}^K\|_{\cH^{1+\delta}_{\xi''}}
\eeqn
for a constant $C_{V,\delta}$ depending only on $V$ and $\delta$, but  
not on $K$ or $N$. 
\end{lemma}

\prf
In a first step, we prove that
\eqn\label{eq-Bdiff-Strich-bd-1}
	\|(B_{N;k+1}^{+}-B_{k+1}^+) \, U^{(k+1)}(t) \gamma_0^{(k+1)}\|_{L^2_{t\in\R}H^1}
	\, \leq \, C \, k^2 \, N^{-\delta\beta} \, \|\gamma_0^{(k+1)}\|_{L^2_{t\in\R}H^{1+\delta}}
\eeqn
holds, for $\widehat V\in C^\delta\cap L^\infty$ with $\delta>0$. 

To this end, we note that
\eqn
	\widehat \VN(\xi) \, = \, \widehat V(N^{-\beta}\xi)
	\; \; \; \; , \; \; \; \;
	\widehat \VN(0) \, = \, \int V_N(x) dx  \, = \, \int V(x) dx \, = \,  \widehat V(0)
	 \, = \, 1 \,, \; \; \;
\eeqn
and we define
\eqn
	\diffN(\xi) \, := \, \frac{N-k}{N} \widehat \VN(\xi) \, - \, \widehat V(0) \,,
\eeqn
We have
\eqn
	\diffN(q-q') \, = \, \diffN^1(q-q') \, + \, \diffN^2(q-q')
\eeqn
where
\eqn
	\diffN^1(q-q') \, := \, \widehat\VN(q-q')-\widehat\VN(0)
	\; \; \; \; , \; \; \; \; 
	\diffN^2(q-q') \, := \, \frac kN \widehat\VN(q-q') \,.
\eeqn
Clearly, we have that for $\delta>0$ small, $\delta$-Holder continuity of $\widehat V$ implies 
\eqn\label{eq-diffN1-bd-1}
	|\diffN^1(q-q')| 
	&\leq& \|\widehat V\|_{C^\delta} \, N^{-\delta \beta}
	| q-q' |^\delta
	\nonumber\\
	&\leq& \|\widehat V\|_{C^\delta} \, N^{-\delta \beta} \,
	( \, | q|^\delta + |q' |^\delta\, )  \,, 
\eeqn
and
\eqn\label{eq-diffN2-bd-1}
	|\diffN^2(q-q')| &\leq&   \|\widehat V\|_{L^\infty}k N^{-1} \,
	%\nonumber\\
	%& \leq & \|\widehat V\|_{C^\e} \, 
	%\Big( \, \frac{\bra u_1+q-q'\ket}{N^\beta} \, + \,
	%\frac{\bra u_1 \ket}{N^\beta} \Big)^\e\, + \,  \|\widehat V\|_{L^\infty}k N^{3\beta-1} \,.
\eeqn 
is clear.  

Next, we let $(\tau,\uu_k,\uu_k')$, $q$ and $q'$
denote the Fourier conjugate variables corresponding to 
$(t,\ux_k,\ux_k')$, $x_{k+1}$, and $x_{k+1}'$, respectively.
Without any loss of generality, we may assume that $j=1$ in
$B_{N;j;k+1}$ and $B_{j;k+1}$. 
Then, 
abbreviating
\eqn
	\delta(\cdots) \, := \, \delta( \, \tau + (u_1+q-q')^2  
	+ \sum_{j=2}^k u_j^2
	+ q^2 
	- |\uu_k'|^2 - (q')^2   \, )
\eeqn
we find
\eqn
	\lefteqn{
	\Big\| \, S^{(k,1)}(B_{N;1;k+1}-B_{1;k+1})U^{(k+1)}(t)\gamma_{0,N}^{(k+1)} \, 
	\Big\|_{L^2 (\R\times\R^{3k}\times\R^{3k})}^2
	}
	\nonumber\\
	& = &
	\int_{\R}d\tau \int d\uu_k d\uu_k' \prod_{j=1}^k\bra u_j\ket^{2} \bra u_j'\ket^{2}
	\\
	&&\quad 
	\Big( \, \int dq dq' \, 
	\delta(\cdots)
	\, \diffN(q-q') \,
	%\nonumber\\
	%&&\quad\quad\quad\quad
	\widehat\gamma^{(k+1)}(\tau,u_1+q-q',u_2,\dots,u_k,q;\uu_k',q')\, \Big)^2 \,,
	\nonumber
\eeqn
similarly as in \cite{klma,kiscst}.
Using the Schwarz inequality, this is bounded by
\eqn\label{eq-DiffStrichartz-bd-1}
	&\leq&\int_{\R}d\tau \int d\uu_k d\uu_k' \, J(\tau,\uu_k,\uu_k')
	\int dq dq' \,  
	\delta(\cdots)
	\nonumber\\
	&&
	\bra u_1+q-q' \ket^{2} \bra q\ket^{2} 
	\bra q'\ket^{2}
	\prod_{j=2}^k\bra u_j\ket^{2} \prod_{j'=1}^k\bra u_{j'}'\ket^{2}
	\, | \, \diffN(q-q') \,|^2
	\nonumber\\
	&&\quad\quad\quad\quad
	\Big| \, \widehat\gamma^{(k+1)}(\tau,u_1+q-q',u_2,\dots,u_k,q;\uu_k',q') \, \Big|^2
\eeqn
where
\eqn\label{eq-Ialpha-def-1}
	%\lefteqn{
	J(\tau,\uu_k,\uu_k') 
	%}
	%\\
	%&&
	\, := \,
	\int_{\R^3\times\R^3} dq \, dq' \, 
	\frac{
	\delta(\cdots) \, \bra u_1\ket^{2}}
	{\bra u_1+q-q'\ket^{2} \bra q\ket^{2} \bra q'\ket^{2}} \,.
\eeqn
The boundedness of
\eqn\label{eq-IN-def-1}
	C_J \, := \,  
	\Big( \, \sup_{\tau,\uu_k,\uu_k'}J(\tau,\uu_k,\uu_k') \,\Big)^{\frac12}
	\, < \, \infty \,
\eeqn
is proven in \cite{klma} for dimension 3, and in \cite{chpa2,kiscst} for dimension 2.

Using \eqref{eq-diffN1-bd-1} and \eqref{eq-diffN2-bd-1}, we obtain, from
the Schwarz inequality, that
\eqn
	\eqref{eq-DiffStrichartz-bd-1} &\leq&C_{V,J}\int_{\R}d\tau \int d\uu_k d\uu_k' \,  
	\int dq dq' 
	\, \big( \,  N^{-2\delta \beta} ( \, |q|^{2\delta}+|q'|^{2\delta} \, ) \, + \, k^2N^{-2} \, \big)
	\nonumber\\
	&&
	\bra u_1+q-q' \ket^{2} \bra q\ket^{2} 
	\bra q'\ket^{2}
	\prod_{j=2}^k\bra u_j\ket^{2} \prod_{j'=1}^k\bra u_{j'}'\ket^{2}
	\nonumber\\
	&&\quad\quad\quad\quad
	\Big| \, \widehat\gamma^{(k+1)}(\tau,u_1+q-q',u_2,\dots,u_k,q;\uu_k',q') \, \Big|^2
\eeqn
where $C_{V,J}:=C_V C_J$, and $C_V$ is a finite constant depending on $V$.
Hence,
\eqn
	\lefteqn{
	\Big\| \, S^{(k,1)}(B_{N;1;k+1}-B_{1;k+1})U^{(k+1)}(t)\gamma_{0,N}^{(k+1)} \, 
	\Big\|_{L^2 (\R\times\R^{3k}\times\R^{3k})}^2
	}
	\nonumber\\
	&  &
	\hspace{1cm} \leq \, C_{V,J} \,  N^{-2\delta \beta} \,
	\|\gamma_{0,N}^{(k+1)}\|_{H^{1+\delta}}^2
	\, + \, C_{V,J} \,  k^2 \, N^{-2} \,
	\|\gamma_{0,N}^{(k+1)}\|_{H^1}^2
	\nonumber\\
	&  &
	\hspace{1cm} \leq \, C_{V,J} \,  k^2 \, N^{-2\delta \beta} \,
	\|\gamma_{0,N}^{(k+1)}\|_{H^{1+\delta}}^2 
\eeqn
follows, given that $\delta\beta<1$ for $\delta>0$ sufficiently small.

Therefore, we conclude that
\eqn
	\lefteqn{
	\| \, (\opB_N \, - \, \opB) \, U(t) \, \Gamma_0^K \, \|_{L^2_{t\in \R}\cH^1_\xi}
	}
	\nonumber\\
	&=&
	\sum_{k=1}^{K} \, \xi^{k} \,
	\|(B_{N;k+1}^{+}-B_{k+1}^+) \, U^{(k+1)}(t) \gamma_0^{(k+1)}\|_{L^2_{t\in\R}H^1} 
	\nonumber\\ 
	&\leq& 
	C  \,  N^{-\delta \beta}   \, \xi^{-1}
	\sum_{k=1}^{K} \, k^2 \, \xi^{k+1} \,  \| \,  \gamma^{(k+1)}_0 \, \|_{H^{1+\delta}_{k+1}} 
	\nonumber\\
	&\leq&
	C \, N^{-\delta \beta}   \, \Big( \sup_k k^2 \Big(\frac\xi{\xi''}\Big)^k\Big) \, \xi^{-1} \,   
	\|\Gamma_0^K\|_{\cH^{1+\delta}_{\xi''}} 
	\nonumber\\
	&\leq&
	C \, N^{-\delta \beta}   \, \xi^{-1} \,   \|\Gamma_0^K\|_{\cH^{1+\delta}_{\xi''}} \,,
\eeqn
for $\xi<\xi''$.
This proves the Lemma.
\endprf

\begin{lemma}
\label{lm-R1bound-2}
Assume that $V_N(x)=N^{3\beta}V(N^\beta x)$ with $\widehat V\in C^\delta\cap L^\infty$. 
Then,
\eqn
	\lefteqn{
	\Big\| \, \int_0^t \big( \, \opB_N -\opB \big) U(t-s) \opB\Gamma^K (s) \, ds
	\, \Big\|_{L^2_{t\in I}\cH^1_{\xi''}}
	}
	\nonumber\\
	&&
	\hspace{3cm}
	\, < \, C_{V,\delta} \,  \xi^{-1} \, T^{\frac12} \, N^{-\delta \beta}  \, 
	\|\opB\Gamma^K\|_{L^2_{t\in I}\cH^{1+\delta}_{\xi'''}}
\eeqn
where the constant $C_{V,\delta}>0$ depends only on $V$ and $\delta$,
and $\xi''<\eta \, \xi'''$ as in \eqref{eq-xiparm-def-1}.
\end{lemma}

\prf
Using Lemma \ref{lm-R1bound-1},
\eqn
	\lefteqn{
	\Big\| \, \int_0^t \big( \, \opB_N -\opB \big) U(t-s) \opB\Gamma^K (s) \, ds
	\, \Big\|_{L^2_{t\in I}\cH^1_{\xi''}}
	}
	\nonumber\\
	&\leq&
	\int_0^T  \, \Big\| \, \big( \, \opB_N -\opB \big) U(t-s) \opB\Gamma^K (s) \, ds
	\, \Big\|_{L^2_{t\in \R}\cH^{1+\delta}_{\xi''}}
	\nonumber\\
	&\leq&
	C_{V,\delta}  \, \xi^{-1} \,  N^{-\delta \beta}  \,
	\int_0^T  \, \big\| \,  \opB\Gamma^K (s) \, ds
	\, \big\|_{\cH^{1+\delta}_{\xi'''}}
	\nonumber\\
	&< & 
	C_{V,\delta}  \, \xi^{-1} \, T^{\frac12} \, N^{-\delta \beta}  \,
	\|\opB\Gamma^K\|_{L^2_{t\in I}\cH^{1+\delta}_{\xi'''}} \,,
\eeqn 
for $C_{V,\delta} $ as in the previous lemma.
This proves the claim. 
\endprf

%$\;$ \\ 

%\newpage

\section{Comparing the $(K,N)$-BBGKY with the full $N$-BBGKY hierarchy}

In this section, we compare solutions $\Gamma_N^K$ of the $(K,N)$-BBGKY
hierarchy to solutions $\Gamma^{\Phi_N}$ to the full $N$-BBGKY hierarchy obtained 
from $\Phi_N$ which solves the $N$-body
Schr\"odinger equation \eqref{eq-NbodySchrod-1}.

\begin{lemma}
\label{lm-BBGKY-NSchrod-diff-1}
Assume that $N$ is sufficiently large, and that in particular,  
\eqref{eq-KnlogNchoice-0-1} holds. 
Then, there is a finite constant $C(T,\xi)$ independent of $K,N$  
such that the estimate
\eqn
	\lefteqn{
	\|\opB_N\Gamma^K_N \, - \, \opB_NP_{\leq K}\Gamma^{\Phi_N}\|_{L^2_{t\in I}\cH_\xi^1}
	}
	\nonumber\\
	&&\hspace{2cm}\leq \, C(T,\xi) \, (\xi')^K \, K \,
	\|(\opB_N\Gamma^{\Phi_N})^{(K)}\|_{L^2_{t\in I}H^1}
\eeqn
holds, where $(\opB_N\Gamma^{\Phi_N})^{(K)}$ is the $K$-th component of
$\opB_N\Gamma^{\Phi_N}$ (and the only non-vanishing component of
$P_K\opB_N\Gamma^{\Phi_N}$), and $\xi'=\xi/\eta$ with $\eta<1$ as specified in 
Lemma \ref{lm-BGamma-Cauchy-1} in the Appendix.
\end{lemma}

\prf
We have already shown that $\opB_N\Gamma_N^K\in L^2_{t\in I}\cH_\xi^1$.
Moreover, it is easy to see that 
\eqn
	\|\opB_N\Gamma_N^K\|_{ L^2_{t\in I}\cH_\xi^1}
	\, < \, C(N,K,T) \,.
\eeqn
The easiest way to see this is to use the trivial bound $\|V_N\|_{L^\infty}<c(N)$, and 
the fact that $I=[0,T]$ is finite.

Thus, 
\eqn\label{eq-BGamma-diff-aux-1}
	\opB_N\Gamma^K_N \, - \, \opB_N P_{\leq K}\Gamma^{\Phi_N}
	\, \in \, L^2_{t\in I}\cH_\xi^1
\eeqn
follows. 

Next, we observe that
\eqn
	\lefteqn{
	(\opB_N\Gamma^K_N \, - \, \opB_N P_{\leq K}\Gamma^{\Phi_N})(t)
	}
	\nonumber\\
	&=&
	(\opB_N\Gamma^K_N \, - \, P_{\leq K-1}\opB_N\Gamma^{\Phi_N})(t) 
	\label{eq-BGamma-diff-aux-0-1}\\
	&=& \opB_N U(t) \Gamma_N^K(0) 
	\, - \, P_{\leq K-1} \opB_N U(t) \Gamma_N(0)
	\nonumber\\
	&& 
	+ \, i  \, \int_0^t \opB_N U(t-s) \opB_N \Gamma_N^K(s) \, ds
	\, - \, i  \, \int_0^t P_{\leq K-1}  \opB_N U(t-s) \opB_N \Gamma^{\Phi_N}(s) \, ds
	\nonumber\\ 
	&=& (\opB_N P_{\leq K} \, - \, P_{\leq K-1} \opB_N) \, U(t) \Gamma_N(0)  
	\nonumber\\
	&&
	\hspace{1cm}+ \,  i  \,  (\opB_N P_{\leq K} \, - \, P_{\leq K-1} \opB_N) \int_0^t     U(t-s) 
	\opB_N \Gamma^{\Phi_N} (s) \, ds
	\nonumber\\
	&&
	\hspace{1cm}+ \,  i  \, \int_0^t   \opB_N U(t-s) 
	\opB_N \Gamma_N^K(s) \, ds
	\nonumber\\
	&&
	\hspace{1cm}
	- \,  i  \,  \opB_N P_{  \leq K} \int_0^t  U(t-s)   \opB_N \Gamma^{\Phi_N}(s) \, ds
	\nonumber\\ 
	&=& (\opB_N P_{\leq K} \, - \, P_{\leq K-1} \opB_N) \,   \Gamma^{\Phi_N} (t)
	\nonumber\\
	&&
	\hspace{1cm}+ \,  i  \, \int_0^t   \opB_N U(t-s) 
	\big( \opB_N \Gamma_N^K \, -  \, P_{\leq K-1} \opB_N \Gamma^{\Phi_N} \big)(s) \, ds
	\nonumber\\
	&&
	\hspace{1cm}
	+ \,  i   \, (\opB_N P_{\leq K-1} - \opB_N P_{\leq K}) 
	\, \int_0^t  U(t-s) \, \opB_N \, \Gamma^{\Phi_N}(s) \, ds \,, \label{eq-Sec6-seqlast}
\eeqn
where to obtain \eqref{eq-BGamma-diff-aux-0-1}
 we used the fact that 
\eqn \label{eq-BPprop-again} 
	(\opB_N P_{\leq K} \, - \, P_{\leq K-1} \opB_N) \,  \Gamma^{\Phi_N}
	\, = \, 0 \,,
\eeqn
which follows, as \eqref{eq-BPprop}, 
based on the componentwise definition of $\opB_N$
in \eqref{eq-Bmain-def-1} and \eqref{eq-Berror-def-1}.
Now we notice that 
$\opB_N P_{\leq K} - \opB_N P_{\leq K-1} =\opB_N P_{K}$. 
Hence \eqref{eq-Sec6-seqlast} implies that 
\eqn
	\lefteqn{
	(\opB_N\Gamma^K_N \, - \, P_{\leq K-1}\opB_N\Gamma^{\Phi_N})(t)
	}
	\nonumber\\
	&=& (\opB_N P_{\leq K} \, - \, P_{\leq K-1} \opB_N) \,   \Gamma^{\Phi_N} (t)
	\nonumber\\
	&& 
	\, - \,  i   \, \int_0^t   \opB_N  \, U(t-s) \, P_{K}   \opB_N \, \Gamma^{\Phi_N}(s) \, ds
	\nonumber\\
	&&
	\hspace{1cm}+ \,  i  \, \int_0^t   \opB_N U(t-s) 
	\big( \opB_N \Gamma_N^K \, -  \, P_{\leq K-1} \opB_N \Gamma^{\Phi_N} \big)(s) \, ds   \,,
\eeqn 
which thanks to  \eqref{eq-BPprop-again} simplifies to 
\eqn
	\lefteqn{
	(\opB_N\Gamma^K_N \, - \, P_{\leq K-1}\opB_N\Gamma^{\Phi_N})(t)
	}
	\nonumber\\
	&=& 
%	(\opB_N P_{\leq K} \, - \, P_{\leq K-1} \opB_N) \,   \Gamma^{\Phi_N} (t)
%	\nonumber\\
%	&& 
	\, - \,  i   \, \int_0^t   \opB_N  \, U(t-s) \, P_{K}   \opB_N \, \Gamma^{\Phi_N}(s) \, ds
	\nonumber\\
	&&
	\hspace{1cm}+ \,  i  \, \int_0^t   \opB_N U(t-s) 
	\big( \opB_N \Gamma_N^K \, -  \, P_{\leq K-1} \opB_N \Gamma^{\Phi_N} \big)(s) \, ds   \,.
\eeqn 
We observe that the term in parenthesis on the  last line  corresponds to 
\eqref{eq-BGamma-diff-aux-0-1}, which is the same as \eqref{eq-BGamma-diff-aux-1}. 
Given $K$, the condition that $N$ is large enough that 
\eqref{eq-KnlogNchoice-0-1} holds, allows us to apply
Lemma \ref{lm-BGamma-Cauchy-1} with
\eqn
	\widetilde\Theta_N^K(t) \, := \, 
	 \opB_N \Gamma_N^K(t) \, -  \, P_{\leq K-1} \opB_N \Gamma^{\Phi_N}(t)
\eeqn
and
\eqn\label{eq-Xiterm-6-diff-1}
	\Xi_N^K(t) & := & 
%	(\opB_N P_{\leq K} \, - \, P_{\leq K-1} \opB_N) \, \Gamma^{\Phi_N} (t)
%	\nonumber\\
%	&&
%	\hspace{1cm}
	- \,  i  \, \int_0^t  \opB_N U(t-s)  P_{  K} \opB_N \Gamma^{\Phi_N}(s) \, ds 
	\,.
\eeqn
%We have
%\eqn
%	(\opB_N P_{\leq K} \, - \, P_{\leq K-1} \opB_N) \,  \Gamma^{\Phi_N}
%	\, = \, 0 \,,
%\eeqn
%thanks to \eqref{eq-BPprop-again}.
We note that for the integral on the rhs of \eqref{eq-Xiterm-6-diff-1},
\eqn
	\lefteqn{
	\|\int_0^t  \opB_N U(t-s)  P_{  K} \opB_N \Gamma^{\Phi_N}(s) \, ds \|_{L^2_{t\in I}\cH_{\xi}^1}
	}
	\nonumber\\
	&&
	\hspace{1cm}
	\, \leq \, C \, T^{\frac12} \, K 
	\, \| P_{K} \, \opB_N  \Gamma^{\Phi_N} \|_{L^2_{t\in I}\cH_{\xi}^1} \,,
\eeqn	
for a constant $C$ uniformly in $N$ and $K$,
based on similar arguments as in the proof of Lemma \ref{lm-R1bound-2},
and using the Strichartz estimates 
\eqref{eq-StrN-main-int} and \eqref{eq-StrN-error-int}.

Accordingly, Lemma \ref{lm-BGamma-Cauchy-1} implies that for $N\gg K$ sufficiently large,
and in particular satisfying 
\eqref{eq-KnlogNchoice-0-1},
\eqn
	\lefteqn{
	\| \opB_N\Gamma^K_N \, - \, \opB_N P_{\leq K}\Gamma^{\Phi_N} \|_{L^2_{t\in I}\cH_\xi^1}
	}
	\nonumber\\
	&&
	\hspace{2cm}= \,
	\| \opB_N\Gamma^K_N \, - \, P_{\leq K-1}\opB_N\Gamma^{\Phi_N} \|_{L^2_{t\in I}\cH_\xi^1} 
	\nonumber\\
	&&
	\hspace{2cm}\leq \,
	C'(T,\xi) \,
	\| \Xi_N^K  \|_{L^2_{t\in I}\cH_{\xi'}^1} 
	\nonumber\\
	&&
	\hspace{2cm}\leq \,
	C(T,\xi) \, K \,
	\| P_{K} \, \opB_N   \, \Gamma^{\Phi_N}  \|_{L^2_{t\in I}\cH_{\xi'}^1} \,,
\eeqn
where $P_K=P_{\leq K}-P_{\leq K-1}$, and  $\xi'=\xi/\eta$ with
$\eta$ specified in Lemma \ref{lm-BGamma-Cauchy-1}.
This immediately implies the asserted estimate, for $T$ sufficiently small (depending on $K$).
Clearly, 
\eqn
	%\lefteqn{
	\| P_{K} \, \opB_N  \Gamma^{\Phi_N} \|_{L^2_{t\in I}\cH_{\xi'}^1}
	%}
	%\nonumber\\
	%&=&
	\, = \, 
	(\xi')^K \, \|(B_{N}\Gamma^{\Phi_N})^{(K)}\|_{L^2_{t\in I}H^1} \,.
\eeqn
Therefore,
\eqn
	\lefteqn{
	\| \opB_N\Gamma^K_N \, - \, \opB_N P_{\leq K}\Gamma^{\Phi_N} \|_{L^2_{t\in I}\cH_\xi^1}
	}
	\nonumber\\
	&&
	\hspace{2cm}\leq \, C(T,\xi) \,
	(\xi')^K \, K \, \|(B_{N}\Gamma^{\Phi_N})^{(K)}\|_{L^2_{t\in I}H^1} \,,
\eeqn
as claimed. Here, we have used the result of Lemma \ref{lm-BGamma-Cauchy-1},
and used the fact that $P_{K} \, \opB_N  \Gamma^{\Phi_N}$ has a single 
nonzero component.
\endprf

%$\;$ \\

%\newpage

\section{Control of $\Gamma^{\Phi_N}$
and $\Gamma_N^K$ as $N\rightarrow\infty$}

In this section, we control the comparison between $\Gamma^{\Phi_N}$
and $\Gamma_N^K$ in a limit where
simultaneously, $N\rightarrow\infty$ and $K=K(N)\rightarrow\infty$
at a suitable rate.

\begin{proposition}
\label{prp-NBBGKY-BNGammaNK-lim-1}
Writing $\beta=\frac{1-\delta'}4$, let
\eqn
	K(N) & = & \frac{\delta'}{2\log C_0}\log N \,,
\eeqn
and $\xi>0$ small enough that
\eqn
	\xi & < & \frac1{b_1} e^{-\frac 2{\delta'} (1-4\delta')\log C_0 } \,,
\eeqn 
with constants
$b_1$ as in Lemma \ref{lm-BBGKY-NSchrod-diff-2},  
and $C_0$ as in Lemma \ref{lm-boardgame-est-1}. Then,
\eqn
	\lim_{N\rightarrow\infty}\|\opB_N\Gamma^{K(N)}_N 
	\, - \, P_{\leq K(N)-1}\opB_N\Gamma^{\Phi_N}\|_{L^2_{t\in I}\cH_\xi^1}
	\, = \, 0 \,
\eeqn
holds.
\end{proposition}

\prf
From Lemma \ref{lm-BBGKY-NSchrod-diff-2} below, we have the estimate
\eqn
	\lefteqn{
	\|\opB_N\Gamma^K_N \, - \, P_{\leq K-1}\opB_N\Gamma^{\Phi_N}\|_{L^2_{t\in I}\cH_\xi^1}
	}
	\nonumber\\
	&&
	\hspace{2cm} \leq \, C(T,\xi) \, N^{4\beta} \, K^2 \,
	(b_1 \xi)^K \, 
\eeqn
where $b_1$, $C_0$ are independent of $K,N$. 

One can easily check that the stated assumptions on $K(N)$ and $\xi$ imply that  
\eqn
	N^{4\beta} \, K^2 \,  (b_1 \xi)^{K(N)} 
	& < &   N^{-\epsilon} \,,
\eeqn
for some $\epsilon>0$.

We note that the given choice of $K(N)$ complies with  \eqref{eq-KNdef-5-1}
and the hypotheses of Lemma \ref{lm-BGamma-Cauchy-1},
which are needed for results in previous sections.
This immediately implies the claim.
\endprf

\begin{lemma}
\label{lm-BBGKY-NSchrod-diff-2}
The estimate
\eqn
	\lefteqn{
	\|\opB_N\Gamma^K_N \, - \, P_{\leq K-1}\opB_N\Gamma^{\Phi_N}\|_{L^2_{t\in I}\cH_\xi^1}
	}
	\nonumber\\
	&&
	\hspace{2cm} \leq \, C(T,\xi )\,   \, N^{4\beta} \, K^2 \,
	(b_1 \xi)^K \, 
\eeqn
holds for finite constants $b_1$, $C(T,\xi)$ independent of $K$ and $N$.
The constant
$b_1$ only depends on the initial state $\Phi_N(0)$ of the $N$-body Schr\"odinger problem,
and on the constant $\eta$ as defined in Lemma \ref{lm-BGamma-Cauchy-1}.
\end{lemma}

\prf
From Lemma \ref{lm-BBGKY-NSchrod-diff-1}, we have that 
\eqn
	\lefteqn{
	\|\opB_N\Gamma^K_N \, - \, P_{\leq K-1}\opB_N\Gamma^{\Phi_N}\|_{L^2_{t\in I}\cH_\xi^1}
	}
	\nonumber\\
	&&
	\hspace{2cm} \leq \, C(T,\xi) \,
	(\eta^{-1}\xi)^K \, K \, \|(B_{N}\Gamma^{\Phi_N})^{(K)}\|_{L^2_{t\in I}H^1}
\eeqn
holds for a finite constant $C(T,\xi)$ independent of $K$, $N$.

The fact that in dimension $d=3$,
\eqn
	\|\VN \|_{C^1} \, \leq \, C \,  N^{4\beta}
\eeqn
follows immediately from the
definition of $\VN$.
Thus, we have
\eqn
	\lefteqn{
	\|(B_{N}^+\Gamma^{\Phi_N})^{(K)}\|_{L^2_{t\in I}H^1}^2
	}
	\nonumber\\
	&\leq&
	C \, 
	\int_I dt \int d\ux_K d\ux_K'  \, \Big|
	\sum_{\ell=1}^K
	\int \Big[\prod_{j=1}^K\bra\nabla_{x_j}\ket\bra\nabla_{x_j'}\ket\Big]
	V_N(x_\ell-x_{K+1})\Phi_N(t,\ux_N)
	\nonumber\\
	&&
	\hspace{5cm}
	\overline{\Phi_N(t,\ux_K',x_{K+1},\dots,x_N) } \, dx_{K+1}\cdots dx_N
	\Big|^2
	\nonumber\\
	&\leq&
	C \,  \|\VN \|_{C^1}^2 \,
	\int_I dt \int d\ux_N \, \Big|
	\sum_{\ell=1}^K
	\int \Big[ \prod_{j=1}^K\bra\nabla_{x_j}\ket \Big] \,  \Phi_N(t,\ux_N) \, \Big|^2
	\nonumber\\
	&&
	\hspace{3cm}
	\sup_{t\in I}
	\int  d\ux_N' 
	\Big| \Big[ \prod_{j=1}^K \bra\nabla_{x_j'}\ket\Big] \, 
	\overline{\Phi_N(t,\ux'_N) }  
	\Big|^2
	\label{eq-Nbody-est-aux-1}\\
	&\leq &
	C \, T  \, N^{8\beta} \, K^2  
	\sup_{t\in I}
	\Big( \, \tr\big( \, S^{(K,1)} \, \gamma_N^{(K)} \, \big) \, \Big)^2
	\label{eq-Nbody-est-aux-2}
\eeqn
using Cauchy-Schwarz to pass to \eqref{eq-Nbody-est-aux-1},
and admissibility to obtain \eqref{eq-Nbody-est-aux-2}.
We also recall that $\gamma_N^{(K)}$ is positive and self-adjoint.

It remains to bound the term
\eqn
	 \tr\big( \, S^{(K,1)} \, \gamma_N^{(K)} \, \big) 
\eeqn 
in \eqref{eq-Nbody-est-aux-2}. To this end, we recall 
energy conservation in the $N$-body Schr\"odinger equation 
satisfied by $\Phi_N$. Indeed, it is proved in  \cite{esy1,esy2,kiscst} that
\eqn
	\Bra \, \Phi_N \, , \, (N+H_N)^{K} \, \Phi_N \, \Ket
	\, \geq \, C^K \, N^{K} \, \tr(S^{(1,K)}\gamma_{\Phi_N}^{(K)})
\eeqn
for some positive constant $C>0$
where
\eqn
	\gamma_{\Phi_N}^{(k)}
	\, = \, \tr_{k+1,\dots,N}(|\Phi_N\rangle\langle\Phi_N|) \,.
\eeqn
This implies that
\eqn
	\tr\big( \, S^{(K,1)} \, \gamma_N^{(K)} \, \big)  \, < \, (b_1')^K
\eeqn
for some finite constant $b_1'>0$.  We then define $b_1:=b_1'\eta^{-1}$.
\endprf

$\;$ \\

%\newpage

\section{Proof of the main Theorem \ref{thm-main-1}}

We may now collect all estimates proven so far, and prove the main result
of this paper, Theorem \ref{thm-main-1}.

To this end, we recall again the solution $\Gamma^K$ of the GP hierarchy with 
truncated initial data, $\Gamma^K(t=0)=P_{\leq K}\Gamma_0\in\cH_\xi^1$.
In \cite{chpa4}, we proved the existence of a solution  $\Gamma^K$
that satisfies the $K$-truncated GP-hierarchy in  integral form, 
\eqn \label{eq-outl-GP-Duhamel-special-3}
	\Gamma^K(t) \, = \, U(t)\Gamma^K(0) \, + \, i \,  \int_0^t U(t-s) \, \opB\Gamma^K(s) \, ds
\eeqn 
where $(\Gamma^K)^{(k)}(t) = 0$ for all $k > K$. Moreover, 
it is shown in \cite{chpa4} that this solution satisfies
$\opB\Gamma^K\in L^2_{t\in I}\cH^{1}_{\xi}$.

Moreover, we proved in \cite{chpa4}  the following convergence:
\begin{enumerate} 
\item[(a)] The strong limit 
\eqn\label{eq-Gamma-GammaK-1}
	\Gamma \, := \, s-\lim_{K\rightarrow \infty} \Gamma^K 
	\; \; \; \; \in \; L^\infty_t \cH^{1}_{\xi} 
\eeqn
exists. \\

\item[(b)] The strong limit  
\eqn\label{eq-Thetalim-GammaK-1}
	\Theta \, := \, s-\lim_{K \rightarrow \infty} \opB \Gamma^K
	\; \; \; \in \;  L^2_t \cH^{1}_{\xi} \,.
\eeqn 
exists, and in particular,
\eqn\label{eq-Thetalim-GammaK-2}
	\Theta \, = \, \opB\Gamma \,.
\eeqn
\end{enumerate}

Clearly, we have that
\eqn
	\lefteqn{
	\|\opB\Gamma  
	\, - \, \opB_N P_{\leq K(N)} \Gamma^{\Phi_N}\|_{L^2_{t\in I}\cH_\xi^1}
	}
	\nonumber\\
	&\leq&\|\opB\Gamma-\opB\Gamma^{K(N)}\|_{L^2_{t\in I}\cH_\xi^1}
	\label{eq-mainThm-prf-1}\\
	&&\, + \,
	\|\opB\Gamma^{K(N)}-\opB_N\Gamma^{K(N)}_N \|_{L^2_{t\in I}\cH_\xi^1}
	\label{eq-mainThm-prf-2}\\
	&&+\,
	\|\opB_N\Gamma^{K(N)}_N 
	\, - \, \opB_N P_{\leq K(N)} \Gamma^{\Phi_N}\|_{L^2_{t\in I}\cH_\xi^1}
	\label{eq-mainThm-prf-3} \,. 
\eeqn
In the limit $N\rightarrow\infty$, we have that $\eqref{eq-mainThm-prf-1}\rightarrow0$ from 
\eqref{eq-Thetalim-GammaK-1} and \eqref{eq-Thetalim-GammaK-2}.

Moreover, $\eqref{eq-mainThm-prf-2}\rightarrow0$ follows from
Proposition \ref{prp-GammaKN-conv-1}.

Finally, $\eqref{eq-mainThm-prf-3}\rightarrow0$ follows from
Proposition \ref{prp-NBBGKY-BNGammaNK-lim-1}.

Therefore,
\eqn
	\lim_{N\rightarrow\infty}\|\opB\Gamma  
	\, - \,\opB_N  P_{\leq K(N)} \Gamma^{\Phi_N}\|_{L^2_{t\in I}\cH_\xi^1} \, = \, 0 \,
\eeqn
follows.

Moreover, we have that
\eqn
	\lefteqn{
	\|P_{\leq K(N)}\Gamma^{\Phi_N} \, - \, \Gamma\|_{L^\infty_{t\in I}\cH_\xi^1}
	}
	\nonumber\\
	&\leq&
	\|P_{\leq K(N)}\Gamma^{\Phi_N} \, - \, \Gamma^{K(N)}_N\|_{L^\infty_{t\in I}\cH_\xi^1}
	\label{eq-mainThm-prf-4}\\
	&&
	\, + \, 
	\|\Gamma^{K(N)} \, - \, \Gamma\|_{L^\infty_{t\in I}\cH_\xi^1}
	\label{eq-mainThm-prf-5}\\
	&&
	\, + \, 
	\|\Gamma^{K(N)}_N \, - \, \Gamma^{K(N)}\|_{L^\infty_{t\in I}\cH_\xi^1} \,.
	\label{eq-mainThm-prf-6}
\eeqn	
In the limit $N\rightarrow\infty$,
we have $\eqref{eq-mainThm-prf-4}\rightarrow0$, as a consequence of 
Proposition \ref{prp-NBBGKY-BNGammaNK-lim-1}.
Indeed,
\eqn
	\lefteqn{
	\|P_{\leq K(N)}\Gamma^{\Phi_N} \, - \, \Gamma^{K(N)}_N\|_{L^\infty_{t\in I}\cH_\xi^1}
	}
	\nonumber\\
	&&
	\, \leq \, 
	T^{\frac12}
	\|\opB_N\Gamma^{K(N)}_N 
	\, - \, \opB_N P_{\leq K(N)} \Gamma^{\Phi_N}\|_{L^2_{t\in I}\cH_\xi^1}
\eeqn
where the rhs tends to zero as $N\rightarrow\infty$, as discussed for \eqref{eq-mainThm-prf-3}.

Moreover,  $\eqref{eq-mainThm-prf-5}\rightarrow0$, as a consequence of 
\eqref{eq-Gamma-GammaK-1}.

Finally, $\eqref{eq-mainThm-prf-6}\rightarrow0$ follows from 
Proposition \ref{prp-GammaKN-conv-1}.

This completes the proof of Theorem \ref{thm-main-1}.
\qed

\newpage

\appendix

\section{Strichartz estimates for GP and BBGKY hierarchies} 

In Appendices A and B, we will prove certain technical results, which we
formulate for dimensions $d\geq1$, respectively $d\geq2$.

In Appendix A, motivated by the Strichartz estimate for the GP hierarchy, 
we establish a Strichartz estimate  for the BBGKY hierarchy. 

\subsection{Strichartz estimates for the GP hierarchy} 

Following \cite{chpa4},  
we first recall a version of the GP Strichartz estimate for the free evolution
$U(t)=e^{it\opDelta_\pm}=(U^{(n)}(t))_{n\in\N}$. 
The estimate is obtained via
reformulating the Strichartz estimate proven by Klainerman and Machedon in \cite{klma}.

\begin{lemma}\label{lm-global-free-Strichartz-1}
Let 
\eqn\label{eq-alphasetcubic-def-1}
	\alpha \, \in \, \alphaset(d)
	\, = \,  \left\{
	\begin{array}{cc}
	(\frac12,\infty) & {\rm if} \; d=1 \\ 
	(\frac {d-1}2 , \infty) & {\rm if} \; d\geq2 \; {\rm and} \; d\neq 3\\
	\big[1,\infty) & {\rm if} \; d \, = \, 3 \,.
	\end{array}
	\right.
\eeqn    
Then, the following hold:
\begin{enumerate}
\item
\underline{Bound for  $K$-truncated case:} 
Assume that $\Gamma_0\in\cH_{\xi}^\alpha$ for some $0 < \xi<1$.  
Then, for any $K\in\N$, there exists a constant $C(K)$ such that 
the Strichartz estimate for the free evolution 
\eqn\label{eq-freeStrichartz-L2H-2}
	\|\opB U(t) \Gamma_0^K\|_{L_{t\in \R}^2\cH_{\xi }^\alpha} \,
	\leq \, \xi^{-1} \, C( K) \,   \|\Gamma_0^K\|_{\cH_{\xi}^\alpha} \, 
\eeqn  
holds. Notably, the value of $\xi$ is the same on both the lhs and rhs.
\\ 

\item
\underline{Bound for $K\rightarrow\infty$:}
Assume that $\Gamma_0\in\cH_{\xi'}^\alpha$ for some $0 < \xi' <1$.  
Then, for any $0<\xi < \xi'$, there exists a constant $C(\xi,\xi')$ such that 
the Strichartz estimate for the free evolution 
\eqn\label{eq-freeStrichartz-L2H-3}
	\|\opB U(t)\Gamma_0\|_{L_{t\in \R}^2\cH_{\xi }^\alpha} \,
	\leq \, C(\xi, \xi') \,  
	\|\Gamma_0 \|_{\cH_{\xi' }^\alpha} \,   
\eeqn  
holds. 
\end{enumerate}
\end{lemma}

\prf 
From Theorem 1.3 in \cite{klma} 
%and Proposition A.1 in \cite{chpa2}, 
we have, for $\alpha\in\alphaset(d,p)$, that
\eqn
	\lefteqn{
	\|B_{k+1}U^{(k+1)}(t)\gamma_0^{(k+1)}\|_{L^2_{t\in\R} H^\alpha_k}
	}
	\nonumber\\
	& \leq &2 \,
	\sum_{j=1}^k \, \| \, B_{j;k+1}^+  
	\, U^{(k+1}(t)\gamma_0^{(k+1)} \, \|_{L^2_{t\in\R} H^\alpha_k}
	\nonumber\\
	& \leq & C \, k \,  \| \, \gamma_0^{(k+1)} \, \|_{H^\alpha_{k+\1}} \,. 
	\label{eq-freeStrichartz-L2H-1-old}
\eeqn  
Then for any  $0 < \xi < \xi'$, we have: 
\begin{align}
	\|\opB U(t)\Gamma_0\|_{ L^2_{t\in\R}\cH_{\xi}^{\alpha} }
	& \leq 
	\sum_{k\geq1} \xi^k \|B_{k+1}U^{(k+1)}(t)
	\gamma_0^{(k+1)}\|_{L^2_{t\in\R} H^\alpha_k}
	\nonumber\\
	& \leq 
	C \, \sum_{k\geq1} k \, \xi^k \, \| \gamma_0^{(k+1)}\|_{H^\alpha_{k+1} } 
	\label{eq-Strprf-usefree}\\
	& = C \, ({\xi'})^{-1} \, \sum_{k\geq1} k \, \left( \frac{\xi}{\xi'} \right)^k \, 
	({\xi'})^{(k+1)} \, \| \gamma_0^{(k+1)}\|_{H^\alpha_{k+1}} 
	\nonumber \\
	& \leq C \, ({\xi'})^{-1} \,  \sup_{k \geq 1} k \left( \frac{\xi}{\xi'}\right)^k \, \sum_{k\geq1} \, 
	({\xi'})^{(k+1)} \, \| \gamma_0^{(k+1)}\|_{H^\alpha_{k+1}} 
	\nonumber \\
	& \leq C(\xi, \xi') \, \|\Gamma_0\|_{\cH_{\xi'}^\alpha} \, ,
	\nonumber 
\end{align} 
where we used \eqref{eq-freeStrichartz-L2H-1-old} to obtain \eqref{eq-Strprf-usefree}.

On the other hand, we have
\begin{align}
	\|\opB U(t)\Gamma_0^K\|_{ L^2_{t\in\R}\cH_{\xi}^{\alpha} }
	& \leq 
	\sum_{k=1}^{K-1} \xi^k \|B_{k+1}U^{(k+1)}(t)
	\gamma_0^{(k+1)}\|_{L^2_{t\in\R} H^\alpha_k}
	\nonumber\\
	& \leq 
	C' \, \sum_{k=1}^{K-1} k \, \xi^k \, \| \gamma_0^{(k+1)}\|_{H^\alpha_{k+1} } 
	\label{eq-Strprf-usefree-1}\\
	& = C' \, K \, ({\xi'})^{-1} \, \sum_{k=1}^{K-1}   \, 
	\xi^{(k+1)} \, \| \gamma_0^{(k+1)}\|_{H^\alpha_{k+1}} 
	\nonumber \\ 
	& \leq C(K) \, \xi^{-1} \, \|\Gamma_0^K\|_{\cH_{\xi}^\alpha} \, .
	\nonumber 
\end{align}
This proves the Lemma.
\endprf

\newpage

\subsection{Strichartz estimates for the BBGKY hierarchy}

In this subsection, we prove a new Strichartz estimate  
for the free evolution 
$U(t)=e^{it\Delta^\pm}$ in $L^2_{t\in I}\cH_\xi^\alpha$,
for the BBGKY hierarchy, at the level of finite $N$.
This result parallels the one for the GP hierarchy, which was stated in 
Lemma \ref{lm-global-free-Strichartz-1}. 

\begin{proposition}
\label{prp-KStrichartz-1}
Let $\alpha \in \alphaset(d)$ for $d\geq2$, and
\eqn\label{eq-betasmall-cond-1}
	\beta \, < \, \frac{1}{ d+2\alpha-1} \,.
\eeqn
Assume that $V  \in L^1(\R^d)$, and that $\widehat V$ decays rapidly outside the unit ball.
Letting $c_0:=1-\beta(d+2\alpha-1)$, the following hold:
\begin{enumerate}
\item
\underline{Bound for  $K$-truncated case:} 
Assume that $\Gamma_0\in\cH_{\xi}^\alpha$ for some $0 < \xi<1$.  
Then, for any $K\in\N$, there exists a constant $C(K)$ such that 
the Strichartz estimate for the free evolution 
\eqn\label{eq-freeKStrichartz-1}
	\|\opB_N^{main} U(t)P_{\leq K}\Gamma_0\|_{L_{t\in \R}^2\cH_{\xi }^\alpha} \,
	\leq \, \xi^{-1} \, C \, K \,   \|\Gamma_0\|_{\cH_{\xi}^\alpha} \, 
\eeqn  
and
\eqn\label{eq-freeKStrichartz-2}
	\|\opB_N^{error} U(t)P_{\leq K}\Gamma_0\|_{L_{t\in \R}^2\cH_{\xi }^\alpha} \,
	\leq \, \xi^{-1} \, C \, K^2 \,  N^{-c_0} \|\Gamma_0\|_{\cH_{\xi}^\alpha} \,.
\eeqn   
Notably, the value of $\xi$ is the same on both the lhs and rhs.
\\ 

\item
\underline{Bound for $K\rightarrow\infty$:}
Assume that $\Gamma_N(0)\in\cH_{\xi'}^\alpha$ for some $0 < \xi' <1$.  
Then, for any $0<\xi < \xi'$, there exists a constant $C(\xi,\xi')$ such that 
we have the Strichartz estimates for the free evolution 
\eqn\label{eq-freeStrichartz-L2H-1}
	\|\opBN^{main}\opU(t)\GammaN(0)\|_{L_{t\in \R}^2\cH_{\xi }^\alpha} \,
	\leq \, C(\xi, \xi') \,  
	\|\GammaN(0)\|_{\cH_{\xi' }^\alpha} \,   
\eeqn  
and
\eqn\label{eq-freeStrichartz-L2H-4}
	\|\opBN^{error}\opU(t)\GammaN(0)\|_{L_{t\in \R}^2\cH_{\xi }^\alpha} \,
	\leq \, C(\xi, \xi') \,  N^{-c_0} \,
	\|\GammaN(0)\|_{\cH_{\xi' }^\alpha} \,.
\eeqn   
\end{enumerate}
\end{proposition}

\prf 
We recall that $\opBN$ contains a main, and an error term.
We will see that 
%the main term is uniformly bounded in $N$, for $0<\beta\leq1$, while
the error term is small only if  the condition
\eqref{eq-betasmall-cond-1} on the values of $\beta$ holds.  
This is an artifact of the $L^2$-type norms used in this paper;
squaring the potential $\VN$ in the error term makes it more singular
to a degree that it can only be controlled for sufficiently small $\beta$.
\\

\underline{\em (1) The main term.}
We first consider the main term in $B_{N;k;k+1}^{\pm} \gammaN^{(k+1)}$. 
We have
\eqn
	\lefteqn{
	\|B_{N;j;k+1}^{+,main}U^{(k+1)}(t) \gammaN^{(k+1)}(0)\|_{L^2_{t\in\R}H^\alpha}^2 
	}
	\nonumber\\
	&=& 
	\|B_{N;k;k+1}^{+,main} U^{(k+1)}(t) \gammaN^{(k+1)}(0)\|_{L^2_{t\in\R}H^\alpha}^2  
	\nonumber\\
	&=&
	\int_{\R}dt
	\int d\ux_k d\ux_k' \, 
	\Big| \, S^{(k,\alpha)}\int d\uu_{k+1} d\uu_{k+1}'  \, \int dx_{k+1}
	 \int dq \, \widehat\VN(q) \, e^{iq(x_{k}-x_{k+1})} 
	\nonumber\\
	&&\quad\quad\quad 
	 e^{i\sum_{j=1}^{k} (x_ju_j-x'_ju_j')} \, e^{i x_{k+1}(u_{k+1}-u_{k+1}')} \,
	 \nonumber\\
	 &&\quad\quad\quad\quad\quad\quad  
	 e^{it\sum_{j=1}^{k+1}(u_j^2-(u_j')^2)} 
	\widehat\gammaN^{(k+1)}(0;\uu_{k+1};\uu_{k+1}')\Big|^2
	%\nonumber\\
\eeqn
\eqn	&=&
	\int_{\R}dt
	\int d\ux_k d\ux_k' \, \int  d\uu_{k+1} d\uu_{k+1}'  d\tu_{k+1} d\tu_{k+1}' 
	\, \int dx_{k+1} \, d\tx_{k+1} \, \int dq \, d\tq 
	\nonumber\\
	&&\quad\quad
	\Big[
	\prod_{j=1}^{k-1} \langle u_j \rangle^{\alpha}\langle u_j'\rangle^{\alpha}
	 \langle \tu_j \rangle^{\alpha}\langle \tu_j'\rangle^{\alpha}
	\Big] \,
	%\nonumber\\
	%&&\quad\quad  
	\langle u_k+q\rangle^\alpha \, \langle \tu_k+\tq\rangle^\alpha 
	\, \langle u_k'\rangle^{\alpha} \,  \langle \tu_k'\rangle^{\alpha} \,	
	\nonumber\\
	&&\quad\quad  \quad 
	 \widehat\VN(q) \,  
	 \overline{ \widehat\VN(\tq) }
	 \, \, e^{iq(x_{k}-x_{k+1}) \, - \, i\tq(x_{k}-\tx_{k+1})} 
	 \nonumber\\
	 &&\quad\quad\quad \quad 
	 e^{i\sum_{j=1}^{k} (x_ju_j-x'_ju_j'-x_j\tu_j+x'_j\tu_j')} 
	 %\nonumber\\
	 %&&\quad\quad\quad\quad\quad 
	 \, e^{i x_{k+1}(u_{k+1}-u_{k+1}' ) \, - \,  i\tx_{k+1}(\tu_{k+1} + \tu_{k+1}')}
	 \nonumber\\
	 &&\quad\quad\quad\quad\quad
	 e^{it\sum_{j=1}^{k+1}(u_j^2-(u_j')^2-\tu_j^2+(\tu_j')^2)} 
	 \nonumber\\
	 &&\quad\quad\quad  
	\widehat\gammaN^{(k+1)}(0;\uu_{k+1};\uu_{k+1}') \,
	\overline{\widehat\gammaN^{(k+1)}(0;\tu_{k+1};\tu_{k+1}')} 
	\label{eq-StrBBGKY-main-int1-0-0}\\
%\eeqn
%\eqn
	&=&
	\int d\uu_{k+1} d\uu_{k+1}' \, d\tu_k \, d\tu_{k+1} \,  d\tu_{k+1}'
	\Big[
	\prod_{j=1}^{k-1} \langle u_j \rangle^{2\alpha}\langle u_j'\rangle^{2\alpha}
	\Big]
	\, \langle u_k'\rangle^{2\alpha}
	\nonumber\\
	&&\quad\quad 
	\int dq d\tq \, \, \widehat\VN(q) \, \overline{\widehat\VN(\tq) }\, 
	\langle u_k+q\rangle^\alpha\langle \tu_k+\tq\rangle^\alpha \, 	
	\nonumber\\
	&&\quad\quad\quad 
	\delta(q-\tq+u_k-\tu_k)   \, 
	\delta(-q+u_{k+1}-u_{k+1}')\delta(-\tq+\tu_{k+1}-\tu_{k+1}') \, 
	\nonumber\\
	&&\quad\quad\quad\quad 
	\delta( u_k^2-\tu_k^2+ u_{k+1}^2-\tu_{k+1}^2 -(u_{k+1}')^2+(\tu_{k+1}')^2)
	\label{eq-StrBBGKY-main-int1-0}\\
	&&\quad\quad\quad 
	\widehat\gammaN^{(k+1)}(0;\uu_{k+1};\uu_{k+1}') \,
	\overline{\widehat\gammaN^{(k+1)}(0;\uu_{k-1},\tu_k,\tu_{k+1};\uu_{k}',\tu_{k+1}')}
	\nonumber
\eeqn
To pass from \eqref{eq-StrBBGKY-main-int1-0-0}
to \eqref{eq-StrBBGKY-main-int1-0}, we have first integrated out the variables
$\ux_{k-1},\tux_k$, thus obtaining delta distributions 
$
	\prod_{j=1}^{k-1}\delta(u_j-\tu_j) \, \prod_{\ell=1}^k \delta(u_\ell'-\tu_\ell')
$
enforcing momentum constraints,
which we subsequently eliminate by integrating over the variables $\tu_j$, $\tu_\ell'$,
for $j=1,\dots,k-1$, $\ell=1,\dots,k$.
The first delta distribution in \eqref{eq-StrBBGKY-main-int1-0} stems from integration in
$x_k$, the second and third from integrating in $x_{k+1}$ and $\tx_{k+1}$, and the fourth
from integrating in $t$ (where terms of the form $u_j^2-\tu_j^2$ and
$(u_\ell')^2-(\tu_\ell')^2$ for $j=1,\dots,k-1$, $\ell=1,\dots,k$ 
have canceled, due to the momentum constraints). 
We note that the expression \eqref{eq-StrBBGKY-main-int1-0} 
differs from the corresponding ones in
\cite{chpa,chpa2,klma} where the Fourier transform in $t$ was first taken before
squaring (in particular, the delta implementing energy conservation
in \eqref{eq-StrBBGKY-main-int1-0} is simpler).
Then we have:
\eqn
	&=&
	\int d\uu_{k+1} d\uu_{k}' \, d\tu_k \, d\tu_{k+1}\, 
	\Big[
	\prod_{j=1}^{k-1} \langle u_j\rangle^{2\alpha}\langle u_j'\rangle^{2\alpha}
	\Big]
	\, \langle u_k'\rangle^{2\alpha}
	\label{eq-StrBBGKY-main-int1}\\
	&&
	\int dq d\tq \, \, \widehat\VN(q) \, \overline{\widehat\VN(\tq) }\, 
	\langle u_k+q\rangle^{2\alpha} \, \delta(q-\tq+u_k-\tu_k)   \, 	
	\nonumber\\
	&&\quad
	\delta(u_k^2-\tu_k^2+u_{k+1}^2-(u_{k+1}-q)^2-\tu_{k+1}^2+(\tu_{k+1}-\tq)^2)
	\nonumber\\
	&&\quad\quad
	\widehat\gammaN^{(k+1)}(0;\uu_{k+1};\uu_{k}',u_{k+1}-q) \,
	\overline{\widehat\gammaN^{(k+1)}(0;\uu_{k-1},\tu_k,\tu_{k+1};\uu_{k}',\tu_{k+1}-\tq)}
	\nonumber\\
%\eeqn
%\eqn
	&=&
	\int d\uu_{k+1} d\uu_{k}' \, d\tu_k \, d\tu_{k+1}
	\Big[
	\prod_{j=1}^{k-1} \langle u_j\rangle^{2\alpha}\langle u_j'\rangle^{2\alpha}
	\Big]
	\, \langle u_k'\rangle^{2\alpha}
	\label{eq-StrBBGKY-main-int2}\\
	&&\quad\quad\quad
	\int dq \, \, \widehat\VN(q) \,  \overline{ \widehat\VN(q+u_k-\tu_k) } \, 
	\langle u_k+q\rangle^{2\alpha} \,  
	\nonumber\\
	&&\quad
	\delta(u_k^2-\tu_k^2+
	u_{k+1}^2-(u_{k+1}-q)^2-\tu_{k+1}^2+(\tu_{k+1}-(q+u_k-\tu_k))^2)
	\nonumber\\
	&&
	\widehat\gammaN^{(k+1)}(0;\uu_{k+1};\uu_{k}',u_{k+1}-q) \,
	\nonumber\\
	&&\quad\quad\quad\quad\quad\quad
	\overline{\widehat\gammaN^{(k+1)}(0;\uu_{k-1},\tu_k,\tu_{k+1};\uu_{k}',\tu_{k+1}-(q+u_k-\tu_k))}
	\nonumber
\eeqn
\eqn
	&=&
	\int d\uu_{k+1} d\uu_{k}' \, d\tu_k \, d\tu_{k+1}
	\Big[
	\prod_{j=1}^{k-1} \langle u_j\rangle^{2\alpha}\langle u_j'\rangle^{2\alpha}
	\Big]
	\, \langle u_k'\rangle^{2\alpha}
	\label{eq-StrBBGKY-main-int3}\\
	&&
	\int dq \, \, \widehat\VN(q+\tu_k) \, \overline{ \widehat\VN(q+u_k) } \, 	
	\langle u_k+\tu_k+q\rangle^{2\alpha} \,
	\nonumber\\
	&&\quad
	\delta(u_k^2-\tu_k^2+u_{k+1}^2-(u_{k+1}-
	q-\tu_k)^2-\tu_{k+1}^2+(\tu_{k+1}-q-u_k))^2)
	\nonumber\\
	&& 
	\widehat\gammaN^{(k+1)}(0;\uu_{k+1};\uu_{k}',u_{k+1}-q-\tu_k) \,
	\overline{\widehat\gammaN^{(k+1)}(0;\uu_{k-1},\tu_k,\tu_{k+1};\uu_{k}',\tu_{k+1}-q-u_k)}
         \nonumber
\eeqn
where to obtain \eqref{eq-StrBBGKY-main-int1} we integrated out 
 the variables $u_{k+1}'$, $\tu_{k+1}'$, to obtain  \eqref{eq-StrBBGKY-main-int2}
 we integrated out the variable $\tq$ and to obtain  \eqref{eq-StrBBGKY-main-int3}
we performed the shift $q\rightarrow q+\tu_k$. 
The last expression is manifestly real and non-negative. One immediately finds the upper bound
\eqn
	&\leq&\|\widehat\VN\|_{L^\infty}^2 \,
	\int d\uu_{k+1} d\uu_{k}' \, d\tu_k \, d\tu_{k+1} \, 
	dq \,   
	\langle u_k+\tu_k+q\rangle^{2\alpha} \,
	\Big[
	\prod_{j=1}^{k-1} \langle u_j\rangle^{2\alpha}\langle u_j'\rangle^{2\alpha}
	\Big]
	\, \langle u_k'\rangle^{2\alpha}
	\nonumber\\
	&&\quad
	\delta(u_k^2-\tu_k^2+u_{k+1}^2-(u_{k+1}-q-\tu_k)^2-\tu_{k+1}^2+(\tu_{k+1}-q-u_k))^2)
	\nonumber\\
	&&
	\widehat\gammaN^{(k+1)}(0;\uu_{k+1};\uu_{k}',u_{k+1}-q-\tu_k) \,
	\overline{\widehat\gammaN^{(k+1)}(0;\uu_{k-1},\tu_k,\tu_{k+1};\uu_{k}',\tu_{k+1}-q-u_k)}
	\nonumber\\
	&=&\|\widehat\VN\|_{L^\infty}^2 \, \| \, B^+_{k;k+1}U^{(k+1)}(t)
	\gamma^{(k+1)}_0 \, \|^2_{L^2_{t\in\R}H^{\alpha}_{k}}
	\nonumber\\
	&\leq&C \, \| \,  \gamma^{(k+1)}_0 \, \|^2_{H^{\alpha}_{k+1}}
	\,.
\eeqn
Here, we have used $\|\widehat\VN\|_{L^\infty}\leq \|\VN\|_{L^1_x}=\|V_1\|_{L^1_x}$ uniformly in $N$, and
the Strichartz estimate for the free evolution in the (infinite) GP hierarchy.

Therefore, we conclude that
\eqn
	\lefteqn{
	\|B_{N;k+1}^{\pm,main}U^{(k+1)}(t) \gammaN^{(k+1)}(0)\|_{L^2_{t\in\R}H^\alpha_k} 
	}
	\nonumber\\
	&\leq&\frac{k(N-k)}N
	\sup_j\|B_{N;j;k+1}^{\pm,main}U^{(k+1)}(t) \gammaN^{(k+1)}(0)\|_{L^2_{t\in\R}H^\alpha}
	\nonumber\\ 
	&\leq&
	C \, ( k - \frac{k^2}{N}) \, \|\gammaN^{(k+1)}(0)\|_{H^\alpha} \,. 
	\label{eq-StrN-main-int}
\eeqn
Hence we have that 
\eqn
	\lefteqn{ 
	\sum_{k \geq 1} \xi^k \, 
       \|B_{N;k+1}^{\pm,main}U^{(k+1)}(t) \gammaN^{(k+1)}(0)\|_{L^2_{t\in\R}H^\alpha_k} 
	}
	\nonumber\\
     	& \leq & C \, 
	\sum_{k\geq1} (k - \frac{k^2}{N}) \, \xi^k \,  \|\gamma_N^{(k+1)}(0)\|_{H^\alpha_{k+1}}
	\label{eq-StrprfN-main-usefree}\\
	& = & C \, 
	({\xi'})^{-1} \, \sum_{k\geq1} (k - \frac{k^2}{N}) \, \left( \frac{\xi}{\xi'} \right)^k \, 
	({\xi'})^{(k+1)} \, \| \gamma_N^{(k+1)}(0)\|_{H^\alpha_{k+1}} 
	\nonumber \\
	& \leq &  C \,
	({\xi'})^{-1} \,  \sup_{k \geq 1} \left( (k - \frac{k^2}{N}) \, \left( \frac{\xi}{\xi'}\right)^k\right) \, \sum_{k\geq1} \, 
	({\xi'})^{(k+1)} \, \| \gamma_N^{(k+1)}(0)\|_{H^\alpha_{k+1}} 
	\nonumber \\
	& \leq &
	\, C(\xi, \xi') \, (1 + \frac{1}{N})  
	\, \|\GammaN(0)\|_{\cH_{\xi'}^\alpha} \,,
	\label{eq-StrN-main} 
\eeqn
where to obtain \eqref{eq-StrprfN-main-usefree} we used \eqref{eq-StrN-main-int}.
\\

\underline{\em (2) The error term.}
Next, we consider the error terms  $B_{N;k}^{\pm,error} \gammaN^{(k)}$.
By %admissibility, the partial trace in $x_{k+1}$, $x_{k+1}'$ yields $\gammaN^{(k)}$,
symmetry, we have
\eqn
	\lefteqn{
	\|B_{N;i,j;k}^{+,error}U^{(k)}(t) \gammaN^{(k)}(0)\|_{L^2_{t\in\R}H^\alpha}^2 
	}
	\nonumber\\
	&=& 
	\|B_{N;1,2;k}^{+,error} U^{(k)}(t) \gammaN^{(k)}(0)\|_{L^2_{t\in\R}H^\alpha}^2  
	\nonumber\\
	&=&
	\int_{\R}dt
	\int d\ux_k d\ux_k' \, 
	\Big| \, S^{(k,\alpha)}\int d\uu_{k} d\uu_{k}'  \, 
	 \int dq \, \widehat\VN(q) \, e^{iq(x_1-x_{2})} 
	\nonumber\\
	&&\quad\quad\quad 
	 e^{i\sum_{j=1}^{k} (x_ju_j-x'_ju_j')} %e^{i x_{k+1}(u_{k+1}-u_{k+1}')} 
	 e^{it\sum_{j=1}^{k}(u_j^2-(u_j')^2)} 
	\widehat\gammaN^{(k)}(0;\uu_{k};\uu_{k}')\Big|^2
	\nonumber\\
	&=&
	\int dt\int d\ux_k d\ux_k' \, 
	\Big| \, \int d\uu_{k} d\uu_{k}' \, \int dq \, \widehat\VN(q) \, e^{iq(x_1-x_{2})}  \, 	
	\nonumber\\
	&&\quad\quad\quad
	\langle u_1+q\rangle^\alpha\langle u_2-q\rangle^\alpha
	\langle u_1'\rangle^\alpha\langle u_2'\rangle^\alpha
	\prod_{j=3}^k \langle u_j\rangle^\alpha\langle u_j'\rangle^\alpha
	\nonumber\\
	&&\quad\quad\quad\quad\quad\quad  
	 e^{i\sum_{j=1}^k (x_ju_j-x'_ju_j')} 
	 e^{it\sum_{j=1}^{k}(u_j^2-(u_j')^2)}
	\widehat\gammaN^{(k)}(0;\uu_{k};\uu_{k}')\Big|^2
	\nonumber\\
	&=&
	\int  d\uu_{k} d\uu_{k}' d\tu_1 d\tu_2
	\langle u_1'\rangle^{2\alpha}\langle u_2'\rangle^{2\alpha}
	\prod_{j=3}^k \langle u_j\rangle^{2\alpha}\langle u_j'\rangle^{2\alpha} 
	 \nonumber\\
	&&\quad
	\, \int dq \, d\tq \, \, \widehat\VN(q) \, \overline{\widehat\VN(\tq)} \, 
	 \langle u_1+q\rangle^{\alpha} \, \langle u_2-q\rangle^{\alpha} \,
	 \langle \tu_1+\tq\rangle^{\alpha} \, \langle \tu_2-\tq\rangle^{\alpha}
	 \nonumber\\
	 &&\quad\quad\quad\quad
	\delta(q-\tq+u_1-\tu_1) \, \delta(q-\tq-u_2+\tu_2) \, \delta( \, u_1^2+u_2^2-\tu_1^2-\tu_2^2
	% + \sum_{j=3}^k (u_j^2 - (u_j')^2) 
	\, )
	 \nonumber\\
	&&\quad\quad\quad\quad\quad\quad   
	\widehat\gammaN^{(k)}(0;\uu_k ;\uu_{k}')  \,
	\overline{\widehat\gammaN^{(k)}(0;\tu_1,\tu_2,u_3,\dots, u_k ;\uu_{k}') }
	\nonumber\\
	&=&
	\int  d\uu_{k} d\uu_{k}'  \,
	\langle u_1'\rangle^{2\alpha}\langle u_2'\rangle^{2\alpha}
	\prod_{j=3}^k \langle u_j\rangle^{2\alpha}\langle u_j'\rangle^{2\alpha} 
	 \label{eq-StrBBGKY-error-int1}\\
	&&\quad	
	\, \int dq \, d\tq \, \, \widehat\VN(q) \, \overline{\widehat\VN(\tq)} \, 
	 \langle u_1+q\rangle^{2\alpha} \, \langle u_2-q\rangle^{2\alpha} \, 
	 \nonumber\\
	 &&\quad\quad\quad\quad
	\delta( \, u_1^2+u_2^2-(u_1+q-\tq)^2-(u_2-q+\tq)^2
	% + \sum_{j=3}^k (u_j^2 - (u_j')^2)
	\, )
	 \nonumber\\
	&&\quad\quad\quad\quad\quad\quad   
	\widehat\gammaN^{(k)}(0;\uu_k ;\uu_{k}')  \,
	\overline{\widehat\gammaN^{(k)}(0;u_1+q-\tq ,u_2-q+\tq,u_3,\dots, u_k ;\uu_{k}') }
	\nonumber\\ 
	&=&
	\int  d\uu_{k} d\uu_{k}' 
	\Big[\prod_{j=1}^k \langle u_j\rangle^{2\alpha}\langle u_j'\rangle^{2\alpha} \Big]
	\, \int dq \, d\tq \, \, \widehat\VN(q) \, \overline{\widehat\VN(\tq)} 
	 \label{eq-StrBBGKY-error-int2}\\
	&&\quad\quad\quad\quad 
	\delta( \, (u_1-q)^2+(u_2+q)^2-(u_1-\tq)^2-(u_2+\tq)^2 
	%+ \sum_{j=3}^k (u_j^2 - (u_j')^2) 
	\, )
	 \nonumber\\
	&&\quad\quad\quad\quad\quad\quad   
	\widehat\gammaN^{(k)}(0;u_1-q,u_2+q,u_3,\dots, u_k ;\uu_{k}')  
	\nonumber\\
	&&\quad\quad\quad\quad\quad\quad\quad\quad\quad   
	\overline{\widehat\gammaN^{(k)}(0;u_1-\tq,u_2+\tq,u_3,\dots, u_k ;\uu_{k}') }
	\nonumber \\
	&=&
	\int  d\uu_{k} d\uu_{k}' \int dq \, d\tq \, \, \widehat\VN(q) \, \overline{\widehat\VN(\tq)} \, 
	\delta( \, 2(-u_1+u_2+q+\tq)\cdot(q-\tq) 
	%+ \sum_{j=3}^k (u_j^2 - (u_j')^2) 
	\, )
	\nonumber\\
	&&\quad\quad\quad
	\Big[\prod_{j=1}^k \langle u_j\rangle^{2\alpha}\langle u_j'\rangle^{2\alpha}\Big] \,
	\widehat\gammaN^{(k)}(0;u_1-q,u_2+q,u_3,\dots, u_k ;\uu_{k}')  
	\nonumber\\
	&&\quad\quad\quad\quad\quad\quad\quad\quad\quad   
	\overline{\widehat\gammaN^{(k)}(0;u_1-\tq,u_2+\tq,u_3,\dots, u_k ;\uu_{k}') }
	\nonumber \,,
\eeqn
where to obtain \eqref{eq-StrBBGKY-error-int1} we integrated out the variables $\tu_1$, $\tu_2$ 
and to obtain  \eqref{eq-StrBBGKY-error-int2} we performed the shifts $u_1 \rightarrow u_1 - q$ 
and $u_2 \rightarrow u_2 + q$. Clearly, the last expression is bounded by
\eqn
	\|B_{N;i,j;k}^{+,error}U^{(k)}(t) \gammaN^{(k)}(0)\|_{L^2_{t} H_k^\alpha}^2 &\leq&
	C_V(N) \, \|\gammaN^{(k)}(0)\|_{H^\alpha_k}^2
\eeqn
where
\eqn
	C_V(N) & := & \sup_{u_1,u_2}\int dq \, d\tq \, 
	\delta( \, 2(-u_1+u_2+q+\tq)\cdot(q-\tq)  \, )
	\nonumber\\
	&&\quad\quad\quad
	\frac{\widehat\VN(q) \, \overline{\widehat\VN(\tq)} \, 
	\langle u_1 \rangle^{2\alpha}\langle u_2\rangle^{2\alpha}}
	{\langle u_1-q\rangle^{\alpha}\langle u_2+q\rangle^{\alpha}
	\langle u_1-\tq\rangle^{\alpha}\langle u_2+\tq\rangle^{\alpha}} \,,
\eeqn
and $\|\widehat\VN\|_{L^\infty}\leq\|\VN\|_{L^1}\leq C$, uniformly in $N$.
We may assume that $\supp\{\widehat\VN\}\subset B_{CN^\beta}(0)$, for some constant $C$.
The modifications for $\widehat\VN$ non-vanishing, 
but decaying rapidly outside $B_{CN^\beta}(0)$ are straightforward.

Then,  
\eqn 
	C_V(N)& \leq & 
	\sup_{u_1,u_2 \in\R^d;q,\tq\in B_{CN^\beta}(0)}\Big[\frac{ \, 
	\langle u_1 \rangle^{2\alpha}\langle u_2\rangle^{2\alpha}}
	{\langle u_1-q\rangle^{\alpha}\langle u_2+q\rangle^{\alpha}
	\langle u_1-\tq\rangle^{\alpha}\langle u_2+\tq\rangle^{\alpha}}\Big]
	\nonumber\\
	&&\quad\quad
	\sup_{u_1,u_2}\int_{B_{CN^\beta}(0)\times B_{CN^\beta}(0)} dq \, d\tq \, 
	\delta( \, 2(-u_1+u_2+q+\tq)\cdot(q-\tq) 
	%+ \sum_{j=3}^k (u_j^2 - (u_j')^2) 
	\, )
	\nonumber\\
	& \leq &C (N^\beta)^{4\alpha}  
	\sup_{u}\int_{B_{CN^\beta}(0)\times B_{CN^\beta}(0)} dv_+ \, dv_- \, 
	\delta( \, 2(u+v_+)\cdot v_- 
	%+ \sum_{j=3}^k (u_j^2 - (u_j')^2) 
	\, )
	\nonumber\\
	\label{eq-CVN0const-bdN-aux-2-0}\\
	& \leq &C (N^\beta)^{4\alpha}  
	\sup_{u}\int_{B_{CN^\beta}(0)\times \cD_{N^\beta}} dv_+ \, dv_-^\perp \, 
	\frac{1}{|u+v_+|} 
	\label{eq-CVN0const-bdN-aux-2}\\
	\nonumber\\
	& = &C (N^\beta)^{4\alpha+d-1}  
	%\sup_{u_1,u_2}
	\int_{B_{CN^\beta}(0) }  \,  
	\frac{dv_+}{|v_+|} 
	\label{eq-CVN0const-bdN-aux-2-1}\\
	&\leq&
	C \, (N^\beta)^{4\alpha+2d-2} 
%	\nonumber\\
%	&=& C \, N^{\beta(2d+4\alpha-2)} 
	\,.
	\label{eq-CVN0const-bdN-aux-1}
\eeqn
To pass to \eqref{eq-CVN0const-bdN-aux-2-0}, we used
\eqn
	\sup_{u_1,u_2 \in\R^d;q,\tq\in B_{CN^\beta}(0)}\Big[\frac{ \, 
	\langle u_1 \rangle^{2\alpha}\langle u_2\rangle^{2\alpha}}
	{\langle u_1-q\rangle^{\alpha}\langle u_2+q\rangle^{\alpha}
	\langle u_1-\tq\rangle^{\alpha}\langle u_2+\tq\rangle^{\alpha}}\Big]
	\, < \, C \, (N^\beta)^{4\alpha} 
\eeqn 
where we note that the maximum 
is attained for configurations similar to
$u_1=q=\tq=-u_2$, $|q|=O(N^\beta)$.

Moreover, we introduced $v_\pm:=q\pm \tq$ as new variables.
Passing to  \eqref{eq-CVN0const-bdN-aux-2}, we integrated out the delta distribution
with the component of $v_-$ parallel to $u+v_+$, for fixed $v_+$ and $u:=u_1-u_2$.
Accordingly, we denoted by $v_-^\perp$  the $(d-1)$-dimensional variable in the 
hyperplane 
$$\cP \, := \, \{v \in\R^d \, | \, v\,\perp\, (u+v_+)\,\}$$ 
perpendicular to 
$u+v_+$, for $u,v_+$ fixed. 

The integral in $v_-$ is supported on  the set
$\cD_{N^\beta}$, given by the intersection of a ball of 
radius $O(N^\beta)$ with the hyperplane $\cP$.
The measure of  $\cD_{N^\beta}$ is at most $O((N^\beta)^{d-1})$. 
This is accounted for in passing to \eqref{eq-CVN0const-bdN-aux-2-1}.

The integral in $v_+$ in  \eqref{eq-CVN0const-bdN-aux-2} 
over a ball or radius $O(N^\beta)$ in $\R^d$ yields another factor 
$O((N^\beta)^{d-1})$  in dimensions $d\geq2$. 
To make this evident, we have shifted $v_+\rightarrow v_+ +u$
in \eqref{eq-CVN0const-bdN-aux-2-1}.

Similarly, we can bound 
the term $\|B_{N;k}^{-,error}U^{(k)}(t) \gammaN^{(k)}(0)\|_{L^2_{t\in\R}H^\alpha_k}$.

Thus, we conclude that
\eqn
	\lefteqn{
	\|B_{N;k}^{\pm,error}U^{(k)}(t) \gammaN^{(k)}(0)\|_{L^2_{t\in\R}H^\alpha_k} 
	}
	\nonumber\\
	&\leq&\frac{k(k-1)}N
	\sup_j\|B_{N;i,j;k}^{\pm,error}U^{(k)}(t) \gammaN^{(k)}(0)\|_{L^2_{t\in\R}H^\alpha}
	\nonumber\\ 
	&\leq&
	C \, k(k - 1) \, N^{\beta( d+2\alpha-1)-1} \, \|\gammaN^{(k)}(0)\|_{H^\alpha} \,.
	\label{eq-StrN-error-int} 
\eeqn
We may now complete the proof.
\\

\noindent{$\bullet$}
\underline{\em Bound for $K\rightarrow\infty$:}
We have 
\eqn
	\lefteqn{ 
	\sum_{k \geq 1} \xi^k \, 
       \|B_{N;k}^{\pm,error}U^{(k)}(t) \gammaN^{(k)}(0)\|_{L^2_{t\in\R}H^\alpha_k} 
	}
	\nonumber\\
     	& \leq & C \, N^{\beta( d+2\alpha-1)-1} \, 
	\sum_{k\geq1} k(k - 1) \, \xi^k \,  \|\gamma_N^{(k)}(0)\|_{H^\alpha_{k+1}}
	\label{eq-StrprfN-usefree-error}\\
	& = & C \, N^{\beta( d+2\alpha-1)-1} \,
	\sum_{k\geq1} k(k - 1) \, \left( \frac{\xi}{\xi'} \right)^k \, 
	({\xi'})^{(k+1)} \, \| \gamma_N^{(k)}(0)\|_{H^\alpha_{k+1}} 
	\nonumber \\
	& \leq &  C \, N^{\beta( d+2\alpha-1)-1} \,
	\sup_{k \geq 1} \left(k(k - 1) \left( \frac{\xi}{\xi'}\right)^k\right) \, \sum_{k\geq1} \, 
	({\xi'})^{(k+1)} \, \| \gamma_N^{(k)}(0)\|_{H^\alpha_{k+1}} 
	\nonumber \\
	& \leq &
	\, C(\xi, \xi') \, N^{\beta( d+2\alpha-1)-1}  
	\, \|\GammaN(0)\|_{\cH_{\xi'}^\alpha} \,,
	\label{eq-StrN-error} 
\eeqn
where
we used \eqref{eq-StrN-error-int} to obtain \eqref{eq-StrprfN-usefree-error}.

Summarizing, we combine \eqref{eq-StrN-main} and \eqref{eq-StrN-error} to obtain:
\eqn
	\lefteqn{
	\|\opBN\opU(t)\GammaN(0)\|_{L_{t\in \R}^2\cH_{\xi }^\alpha} \,
	}
	\nonumber\\
	& = &
	\sum_{k \geq 1} \xi^k \, \|
	(\opBN\opU(t)\GammaN(0))^{(k)}
	%B_{N;k+1}^{\pm}U^{(k+1)}(t) \gammaN^{(k+1)}(0)
	\|_{L^2_{t\in\R}H_\xi^\alpha}
	\nonumber \\
       	& \leq &
	\sum_{k\geq 1} \xi^k \, \|B_{N;k+1}^{\pm,main}U^{(k+1)}(t) 
	\gammaN^{(k+1)}(0)\|_{L^2_{t\in\R}H_\xi^\alpha}
       	\nonumber \\
	&& \quad \quad + 
	\sum_{k \geq 1} \xi^k \, \|B_{N;k}^{\pm,error}U^{(k)}(t) 
	\gammaN^{(k)}(0)\|_{L^2_{t\in\R}H_\xi^\alpha}
	\nonumber\\
        	& \leq &
	\, C(\xi, \xi') \, (1 + N^{-1} + N^{\beta( d+2\alpha-1)-1})
	\, \|\GammaN(0)\|_{\cH_\xi^\alpha} \,.
\eeqn

\noindent{$\bullet$}
\underline{\em Bound for finite $K$:}
Replacing the infinite sum over indices $k$ in \eqref{eq-StrN-error}  by a finite sum 
with $1\leq k\leq K$, it is easy
to see that one gets
\eqn
	\lefteqn{
	\|\opBN\opU(t)P_{\leq K }\GammaN(0)\|_{L_{t\in \R}^2\cH_{\xi }^\alpha} \,
	}
	\nonumber\\
	& = &
	\sum_{k = 1}^K 
	\xi^k \, \|
	(\opBN\opU(t)P_{\leq K }\GammaN(0))^{(k)}
	%B_{N;k+1}^{\pm}U^{(k+1)}(t) \gammaN^{(k+1)}(0)
	\|_{L^2_{t\in\R}H_\xi^\alpha}
	\nonumber \\
       	& \leq &
	\sum_{k= 1}^K \xi^k \, \|B_{N;k+1}^{\pm,main}U^{(k+1)}(t) 
	\gammaN^{(k+1)}(0)\|_{L^2_{t\in\R}H_\xi^\alpha}
       	\nonumber \\
	&& \quad \quad + 
	\sum_{k = 1}^K \xi^k \, \|B_{N;k}^{\pm,error}U^{(k)}(t) 
	\gammaN^{(k)}(0)\|_{L^2_{t\in\R}H_\xi^\alpha}
	\nonumber\\
        	& \leq &
	C\, K \, \xi^{-1} \, (1 + N^{-1} + K N^{\beta( d+2\alpha-1)-1})
	\, \|\GammaN(0)\|_{\cH_\xi^\alpha} 
\eeqn
(by setting $\xi=\xi'$, and taking $\sup_{1\leq k\leq K}$ in the
second last line of  \eqref{eq-StrN-error} )).

This concludes the proof.
\endprf

\begin{remark}
We note that the restriction on $\beta$ is due to the error
term in $\opBN$. 
It stems from the fact that since we are using the $L^2$-type $H^\alpha$-norms,
the quantity $\VN$, which is essentially a Dirac function, is 
squared. 
We can only expect to get $\beta=1$ if we use $L^1$-type trace norms
similarly as Erd\"os, Schlein and Yau, \cite{esy1,esy2}.

The main term in $\opBN$, on the other hand, does allow for the entire
range $0<\beta\leq1$. This is because in this term, averaging (integration
over the variable $x_{k+1}$, which is part of the argument of $\VN(x_j-x_{k+1})$)
is performed before squaring, in order to obtain the $H^\alpha$-norm.
\end{remark}

%$\;$ 

%\newpage

\section{Iterated Duhamel formula and boardgame argument}

The goal of Appendix B is to 
prove the main  Lemma \ref{lm-BGamma-Cauchy-1} below,
following our earlier work
\cite{chpa2,chpa4}, where we used analogous 
estimates to prove well-posedness
results for the infinite GP hierarchy. 
The proof is
based on the boardgame strategy introduced in 
\cite{klma} (which is a reformulation of a method introduced in \cite{esy1,esy2}).

\begin{definition}
Let
$\widetilde\Xi=(\widetilde{\Xi^{(k)}})_{n\in\N}$ denote 
a sequence of arbitrary Schwartz class functions 
$\widetilde{\Xi^{(k)}}\in{\mathcal S}(\R\times\R^{kd}\times\R^{kd})$.
Then, we define the associated sequence $\duh_j(\Xi)$ 
of  {\em $j$-th level iterated Duhamel terms} based on $B_N^{main}$, with components given by
\eqn\label{eq-Duh-j-def-1}
	\lefteqn{
	\duh_j(\widetilde{\Xi})^{(k)}(t) 
	}
	\\
	& := & (-i\mu)^j\int_0^t dt_1 \cdots \int_0^{t_{j-1}}dt_j
	B_{N;k+1}^{main}
	e^{i(t-t_1)\Delta_\pm^{(k+1)}}
	B_{N;k+2}^{main}e^{i(t_1-t_2)\Delta_\pm^{(k+2)}}
	\nonumber\\
	&&\quad\quad\quad\quad\quad\quad
	B_{N;k+2}^{main}\cdots 
	\cdots e^{i (t_{j-1}-t_j) \Delta_\pm^{(k+j)}}   
	( \, \widetilde{\Xi} \, )^{(k+j)}(t_j) \,. \;\;
	\nonumber
\eeqn 
for $\mu=\pm1$, with the conventions $t_0:=t$, and 
\eqn
	\duh_0(\widetilde\Xi)^{(k)}(t) \, := \, ( \, \widetilde\Xi \, )^{(k)}(t)
\eeqn
for $j=0$.
\end{definition}

Here, the definition is given for Schwartz class functions, and can be extended
to other spaces by density arguments. The fact that 
$\duh_j(\widetilde\Xi)^{(k)}\in{\mathcal S}(\R\times\R^{kd}\times\R^{kd})$
holds in this situation, for all $k$,
can be easily verified.
Using the boardgame strategy of \cite{klma} (which is a reformulation of a 
combinatorial argument developed in \cite{esy1,esy2}), one obtains:

\begin{lemma}
\label{lm-boardgame-est-1}
Let $\alpha\in\alphaset(d)$. 
Then, for $\widetilde\Xi=(\widetilde{\Xi^{(k)}})_{k\in\N}$ as above,
\eqn\label{eq-BGamma-Duhj-combin-bd-1}
	\lefteqn{
	\| \, \duh_j(\widetilde\Xi)^{(k)}(t) 
	\, \|_{L^2_{t\in I}H^\alpha(\R^{kd}\times\R^{kd})} 
	}
	\\
	&&\hspace{1cm}
	\, \leq \, k \, C_0^k \, (c_0 T)^{\frac {j}2} \|\widetilde{\Xi^{(k+j)}}
	\|_{L^2_{t\in I}H^\alpha(\R^{(k+j)d}\times\R^{(k+j)d})}  \,,
	\nonumber
\eeqn 
where the constants $c_0,C_0$ depend only on $d,p$.
\end{lemma}

In \cite{chpa2,klma},  
Lemma \ref{lm-boardgame-est-1} is proven for the operator $\opB$ instead of $\opB_N^{main}$,
based on the use of Lemma \ref{lm-global-free-Strichartz-1}.
For the case of $\opB_N^{main}$, we invoke Proposition \ref{prp-KStrichartz-1} instead;
the argument then proceeds exactly in the same way.
We remark, however, that we do not know if the boardgame argument can be adapted to the case of
$B_N^{error}$.
\footnote{We thank Xuwen Chen for pointing out to us an issue in this regard in an earlier version of this paper.}

We then consider solutions $\Theta_N^K$ of 
the integral equation  
\eqn\label{eq-tildTheta-eq-1}
	\Theta_N^K(t) & = & \Xi_N^K(t) 
	\, + \, i \int_0^t \opB_N \, U(t-s) \, \Theta_N^K(s) ds
	\nonumber\\
	& = & \widetilde{\Xi_N^K(t)} 
	\, + \, i \int_0^t \opB_N^{main} \, U(t-s) \, \Theta_N^K(s) ds
\eeqn
where $(\Xi_N^K)^{(k)}(t)=0$ and $(\Theta^K_N)^{(k)}(t)=0$ for all $k>K$, and all $t\in I=[0,T]$.
Moreover,
\eqn\label{eq-tildTheta-eq-1-1}
	\widetilde{\Xi_N^K(t)} \, := \, \Xi^K(t) \, + \, 
	i\int_0^t {\opB_N^{error}} \, U(t-s) \, \Theta_N^K(s) ds \,.
\eeqn
By iteration of the Duhamel formula,
\eqn
	(\Theta_N^K)^{(k)}(t) & = & \sum_{j=0}^{\ell-1} 
	\duh_j(\widetilde{\Xi_N^K})^{(k)}(t)
	%\nonumber\\
	%&&\hspace{1cm}
	\, + \,  \duh_{k}(\Theta_N^K)^{(k)}(t) \,,
\eeqn
obtained from iterating the Duhamel formula $\ell$ times for the $k$-th component of 
$\Theta_N^K$.
Since $(\Theta_N^K)^{(m)}(t)=0$ for all $m>K$, the remainder term on the 
rhs is zero
whenever $n+\ell>K$. Thus,
 \eqn
	(\Theta_N^K)^{(k)}(t) & = & \sum_{j=0}^{N-k} 
	\duh_j(\widetilde{\Xi_N^K})^{(k)}(t)  \,,
\eeqn
where each term on the right explicitly depends only on  $\widetilde{\Xi_N^K}(t)$
(there is no implicit dependence on the solution $\Theta_N^K(t)$).

\begin{lemma}
\label{lm-BGamma-Cauchy-1}
Assume that $N$ is sufficiently large, and that in particular, defining $\delta'>0$ by
\eqn
	\beta & = & \frac{1-\delta'}{d+2\alpha-1}  \, ,
\eeqn
the condition
\eqn
	K & < & \frac{\delta'}{\log C_0} \, \log N \,,
\eeqn
holds, where the constant $C_0$ is as in Lemma \ref{lm-boardgame-est-1}.

Let $\Theta_N^K$ and $\Xi_N^K$ satisfy \eqref{eq-tildTheta-eq-1}.
%with only the first $K$ components nonzero,  so that 
%$(\widetilde{\opBN}\Gamma)^{(k)}=(\opBN\Gamma)^{(k)}$ for $1\leq k\leq K$.
Assume that  
$\Xi_N^K\in L^2_{t\in I}\cH_{\xi'}^\alpha$ for some $0<\xi'<1$, 
and that $\xi$ is small enough that $0<\xi<\eta\xi'$, with $\eta$ specified in 
\eqref{eq-eta-ineq-def-1}.
Then, the estimate
\eqn 
	\| \Theta_N^K\|_{L^2_{t\in I}\cH_\xi^\alpha}
	%& \leq & C(\xi,\xi') \, \|\Gamma_{N_1}(0)-\Gamma_{N_2}(0)\|_{\cH_{\xi'}^\alpha}
	%\nonumber\\
	%&\leq&
	\, \leq \, C_1(T,\xi,\xi') \,  \|\Xi_N^K\|_{L^2_{t\in I}\cH_{\xi'}^\alpha}
	\label{eq-Gammadiff-Cauchy-aux-2}
\eeqn
holds for a finite constant $C_1(T,\xi,\xi')>0$ independent of $K,N$. 
\end{lemma}

\prf 
We have
 \eqn\label{eq-BGammaN1minN2-1-1}
	(\Theta_N^K)^{(k)}(t) & = & \sum_{j=0}^{ N-k} 
	\duh_j(\widetilde{\Xi_N^K})^{(k)}(t)  \,,
\eeqn
using the fact that $(\Theta_{N}^K)^{(k+j)}=0$ for $j>N-k$, see \eqref{eq-Duh-j-def-1}.
 
Using  Lemma \ref{lm-boardgame-est-1}, we therefore find that
\eqn  
	\lefteqn{
	\|(\Theta_N^K)^{(k)} \|_{L^2_{t\in I}H^\alpha}
	}
	\nonumber\\
	& \leq &   \sum_{j=0}^{ N-k}
	\|(\duh_j(\widetilde{\Xi_N^K})^{(k+1)}(t)\|_{L^2_{t\in I}H^\alpha}
	\nonumber\\
	& \leq &   \sum_{j=0}^{ N-k}
	k \, C_0^k \, (c_0T)^{\frac j2} \, 
	\|(\widetilde{\Xi_N^K})^{(k+j)}\|_{L^2_{t\in I}H^\alpha}
	\nonumber\\
	%& \leq & ( \xi)^{-k} 
	%k C_0^k \, (\xi/\xi')^k \sum_{j=0}^{ N-k} 
	%(c_0T(\xi')^{-2})^{\frac j2}(\xi')^{k+j} 
	%\| (\widetilde{\Xi_N^K})^{(k+j)}\|_{ L^2_{t\in I} H^\alpha}
	%\nonumber\\
	& \leq &(I)_k \, + \, (II)_k
\eeqn
where
\eqn
	(I)_k &:=&  \xi^{-k} 
	k \, C_0^k \, (\xi/\xi')^k
	\sum_{j=0}^{ N-k} 
	(c_0T(\xi')^{-2})^{\frac j2}(\xi')^{k+j}  
	\| (\Xi_N^K)^{(k+j)}\|_{ L^2_{t\in I} H^\alpha}
	\\
	(II)_k &:=& \xi^{-k} 
	k C_0^k \, 
	\sum_{j=0}^{ N-k} 
	(c_0T\xi^{-2})^{\frac j2}\xi^{k+j}  
	\Big\| \int_0^t ({B^{error}_N}U(t-s){\Theta_N^K})^{(k+j)}ds
	\Big\|_{ L^2_{t\in I} H^\alpha} \, ,
	\nonumber
\eeqn
recalling \eqref{eq-tildTheta-eq-1-1}.

We have
\eqn
	(I)_k & \leq &  (\xi)^{-k} 
	k \, C_0^k \, (\xi/\xi')^k \, C(T,\xi') \, 
	\| \Xi_N^K \|_{L^2_{t\in I}  \cH_{\xi'}^\alpha}
	 \,,
	\label{eq-BGammaN1minN2-2}
\eeqn
for $T>0$ sufficiently small so that $c_0T(\xi')^{-2}\leq1$.
Hence, 
\eqn  
	\sum_{k\in\N}\xi^k(I)_k & \leq &     C(T,\xi') 
	\Big(\sum_{k\in\N} k \, C_0^k \, (\xi/\xi')^k \Big)
	\,  \| \Xi_N^K \|_{L^2_{t\in I} \cH_{\xi'}^\alpha}
	\nonumber\\
	& \leq & C'(T,\xi,\xi') \,    \| \Xi_N^K \|_{L^2_{t\in I} \cH_{\xi'}^\alpha}
	 \,,
	\label{eq-BGammaN1minN2-3}
\eeqn
for $\xi<\eta\xi'$ where 
\eqn\label{eq-eta-ineq-def-1}
	\eta \, < \, (\max\{1,C_0\})^{-1}
\eeqn
noting that $C_0=C_0(d,p)$.

To bound $(II)_k$, we note that
\eqn
	\lefteqn{
	\Big\|\int_0^t 
	({\opB_N^{error}} \, U(t-s) \, \Theta_N^K(s) )^{(k+j)} ds
	\Big\|_{L^2_{t\in I}H^\alpha}
	}
	\nonumber\\
	& \leq & 
	\Big\|\int_0^t \|({\opB_N^{error}} \, U(t-s) \, 
	\Theta_N^K)^{(k+j)}(s) \|_{H^\alpha} ds
	\Big\|_{L^2_{t\in I}}
	\nonumber\\
	& \leq & 
	\int_0^T
	\| ({\opB_N^{error}} \, U(t-s) \, 
	\Theta_N^K)^{(k+j)}(s) \|_{L^2_{t\in I}H^\alpha} ds 
	\nonumber\\
	& \leq & 
	C (k+j)^2 N^{\beta(d+2\alpha-1)-1}\int_0^T
	\| (\Theta_N^K)^{(k+j)} \|_{ H^\alpha} ds 
	\nonumber\\
	& \leq & 
	T^{\frac12} \, C (k+j)^2 \,  N^{\beta(d+2\alpha-1)-1} \,
	\| (\Theta_N^K)^{(k+j)} \|_{L^2_{t\in I}H^\alpha}  
\eeqn
using \eqref{eq-StrN-error-int} to pass from the third to the fourth line.
This implies that for $T>0$ small enough that  $c_0T\xi^{-2}\leq1$,
\eqn\label{eq-BGammaN1minN2-3-2}
	\lefteqn{
	\sum_{k=1}^K \xi^k (II)_k
	}
	\nonumber\\ 
	& \leq &  
	T^{\frac12} \, K CC_0^K  \, N^{\beta(d+2\alpha-1)-1} 
	\sum_{j=0}^{ N-k} 
	(c_0T \xi^{-2})^{\frac j2}\xi^{k+j}  (k+j)^2
	\| (\Theta_N^K)^{(k+j)} \|_{L^2_{t\in I}H^\alpha} \, 
	\nonumber\\
	& \leq &  
	T^{\frac12} \, K^3  C C_0^K \, N^{\beta(d+2\alpha-1)-1} 
	\sum_{j=1}^{ K-k} \xi^{k+j}   
	\| (\Theta_N^K)^{(k+j)} \|_{L^2_{t\in I}H^\alpha}
	\nonumber\\
	& \leq &  
	T^{\frac12} \, K^3 C C_0^K \, N^{\beta(d+2\alpha-1)-1}  	\| \Theta_N^K \|_{L^2_{t\in I}\cH^\alpha_\xi} \,.
\eeqn
Here, we used that $(\Theta_N^K)^{(k+j)}=0$ for $k+j > K$ to pass to the
third line.

Summarizing, 
\eqn  
	\| \Theta_N^K \|_{L^2_{t\in I}\cH^\alpha_\xi}
	&=&
	\sum_{k\in\N}\xi^k\|({\Theta^K_N})^{(k)} \|_{L^2_{t\in I}H^\alpha}
	\nonumber\\
	 & \leq & 
	 \sum_{k\in\N}\xi^k \Big( \, (I)_k \, + \, (II)_k \, \Big)
	\nonumber\\
	 & \leq & 
	 C_{N,K}(T,\xi,\xi') \,    \| \Xi_N^K \|_{L^2_{t\in I} \cH_{\xi'}^\alpha} \,
	 \,,
	\label{eq-BGammaN1minN2-3-1}
\eeqn
using \eqref{eq-BGammaN1minN2-3} and \eqref {eq-BGammaN1minN2-3-2}, where
\eqn
	C_{N,K}(T,\xi,\xi') \, := \, 
	\frac{C'(T,\xi,\xi')}{1-T^{\frac12}\,K^3 C C_0^K \,N^{\beta(d+2\alpha-1)-1} }\,.
\eeqn
Letting
\eqn\label{eq-KlogNchoice-1}
	\beta & = & \frac{1-\delta'}{d+2\alpha-1} \; \; , \; \; \delta' \, > \, 0 \, ,
	\nonumber\\
	K & < & \frac{\delta'}{\log C_0} \, \log N \,,
\eeqn
we find that, writing $K= \frac{\delta'-\epsilon}{\log C_0}\log N$ for $\epsilon>0$,
\eqn
	T^{\frac12}\,K^3 C C_0^K \,N^{\beta(d+2\alpha-1)-1}
	\, < \,
	C''(T,\xi)\,(\log N)^3  \,N^{-\epsilon} \, < \, \frac12\,
\eeqn 
for sufficently large $N$, where the constant $C''(T,\xi)$ is independent of $N$. 

We conclude that given \eqref{eq-KlogNchoice-1}, and all $N$ sufficiently large, we have that
\eqn
	C_{N,K}(T,\xi,\xi') \, < \, 2C'(T,\xi,\xi') \,.
\eeqn
This proves the claim.
\endprf

\subsection*{Acknowledgements} 
We thank B. Schlein and H.-T. Yau for helpful comments. 
We also thank X. Chen, J. Colliander, M. Grillakis, S. Klainerman, I. Rodnianski, M. Weinstein
for inspiring discussions.
We are grateful to W. Beckner, A. Figalli, and K. Taliaferro for useful comments.
We thank X. Chen, M. Machedon, and an anonymous referee for very detailed and helpful comments.
The work of T.C. was supported by NSF grants DMS-0940145, DMS-1009448,
and DMS-1151414 (CAREER).
The work of N.P. was supported by NSF grants DMS-0758247 and DMS-1101192
and an Alfred P. Sloan Research Fellowship.

%\newpage

\end{document}